\pdfoutput=1
\RequirePackage{fix-cm}
\documentclass[smallextended]{svjour3}       
\smartqed  
\usepackage{graphicx}
\usepackage{mathptmx}      
\journalname{Experimental Astronomy}
\usepackage{natbib}     \bibpunct{[}{]}{;}{n}{}{,}
\newcommand{\compass}{{\sc{Compass}}}
\newcommand{\solmex}{SolmeX}
\newcommand{\vircor}{VIRCOR}
\newcommand{\cusp}{CUSP}
\newcommand{\susp}{SUSP}
\newcommand{\eip}{EIP}
\newcommand{\chrome}{ChroME}
\newcommand{\arcsec}{$^{\prime\prime}$}
\newcounter{IonCS}
\newcommand{\ion}[2]{\setcounter{IonCS}{#2}#1\,{\small{\Roman{IonCS}}}}
\newcommand{\sect}[1]{Sect.\,\ref{#1}}
\newcommand{\sects}[1]{Sects.\,\ref{#1}}
\newcommand{\fig}[1]{Fig.\,\ref{#1}}
\newcommand{\figs}[1]{Figs.\,\ref{#1}}
\newcommand{\tab}[1]{Table\,\ref{#1}}
\newcommand{\Lya}{Ly-$\alpha$}
\newcommand{\mgiik}{\ion{Mg}{2}\,k}
\newcommand{\pol}{{\mbox{\sc{pol}}}}
\newcommand{\nop}{{\mbox{\sc{n-p}}}}
\usepackage{color}
\newcommand{\nnn}[1]{{#1}}

\begin{document}

\title{Solar magnetism eXplorer (SolmeX)%
}
\subtitle{Exploring the magnetic field in the upper atmosphere\\of our closest star}


\author{H.~Peter        \and 
        L.~Abbo         \and V.~Andretta   \and F.~Auch\`ere      \and A.~Bemporad    \and 
        F.~Berrilli     \and V.~Bommier    \and A.~Braukhane      \and R.~Casini      \and 
        W.~Curdt        \and J.~Davila     \and H.~Dittus         \and S.~Fineschi    \and 
        A.~Fludra       \and A.~Gandorfer  \and D.~Griffin        \and B.~Inhester    \and
        A.~Lagg         \and E.~Landi Degl'Innocenti              \and V.~Maiwald     \and 
        R.~Manso Sainz  \and V.~Mart\'inez Pillet                   \and S.~Matthews    \and 
        D.~Moses        \and S.~Parenti    \and A.~Pietarila      \and D.~Quantius    \and
        N.-E.~Raouafi   \and J.~Raymond    \and P.~Rochus         \and O.~Romberg     \and 
        M.~Schlotterer  \and U.~Sch\"uhle  \and S.~Solanki        \and D.~Spadaro     \and 
        L.~Teriaca      \and S.~Tomczyk    \and J.~Trujillo Bueno \and J.-C.~Vial}


\institute{H.~Peter, W.~Curdt, A.~Gandorfer, B.~Inhester, 
           A.~Lagg, U.~Sch\"uhle, S.~Solanki, L.~Teriaca, \at
              Max Planck Institute for Solar System Research,
              37191 Katlenburg-Lindau, Germany\\
              \email{peter@mps.mpg.de}
           \and
           L.~Abbo, A.~Bemporad, S.~Fineschi,   \at
           INAF Osservatorio Astronomico di Torino, Italy
           \and
           V.~Andretta,   \at
           INAF Osservatorio Astronomico di Capodimonte, Napoli, Italy
           \and
           F.~Auch\`ere, J.-C.~Vial,   \at
           Institut d'Astrophysique Spatiale, Orsay, France
           \and
           F.~Berrilli,   \at
           Universit\`a degli Studi di Roma ``Tor Vergata'', Italy
           \and
           V.~Bommier,   \at
           Observatoire de Paris-Meudon, France
           \and
           A.~Braukhane, H.~Dittus, V.~Maiwald, D.~Quantius, O.~Romberg,
           M.~Schlotterer,   \at
           DLR Institute of Space Systems, Bremen, Germany
           \and
           R.~Casini, S.~Tomczyk,   \at
           NCAR / High Altitude Observatory, Boulder, CO, USA
           \and
           J.~Davila,   \at
           NASA / GSFC, Greenbelt, MD, USA
           \and
           A.~Fludra, D.~Griffin,   \at
           STFC Rutherford Appleton Laboratory, Oxon, UK
           \and
           E.~Landi Degl'Innocenti,   \at
           Universit\`a degli Studi di Firenze, Italy
           \and
           R.~Manso Sainz, V.~Mart\'inez Pillet, J.~Trujillo Bueno,   \at
           Instituto de Astrof\'isica de Canarias, Tenerife, Spain
           \and
           S.~Matthews,   \at
           Mullard Space Science Laboratory, Surrey, UK
           \and
           D.~Moses,   \at
           Naval Research Laboratory, Washington, DC, USA
           \and
           S.~Parenti,   \at
           Royal Observatory of Belgium, Brussels, Belgium
           \and
           A.~Pietarila,   \at
           National Solar Observatory, Tucson, AZ, USA
           \and
           N.-E.~Raouafi,   \at
           Johns Hopkins University / APL, Laurel, USA
           \and
           J.~Raymond,   \at
           Smithsonian Astrophysical Observatory, Cambridge, USA
           \and
           P.~Rochus,   \at
           Centre Spatial de Li\`ege, Universit\'e de Li\`ege, Belgium
           \and
           D.~Spadaro,   \at
           INAF Osservatorio Astrofisico di Catania, Italy
}

\date{\hspace*{\fill}{\sf Version: August 26, 2011}}

\maketitle

\begin{abstract}

\nnn{The magnetic field plays a pivotal role in many fields of Astrophysics. This is especially true} for the physics of the solar atmosphere. \nnn{Measuring the magnetic field in the upper solar atmosphere is crucial to understand} the nature of the underlying physical processes that drive the violent dynamics of the solar corona -- that can also affect life on Earth.

\nnn{\solmex, a fully} equipped solar space observatory for remote-sensing observations, will provide the first comprehensive measurements of the strength and direction of the magnetic field in the upper solar atmosphere. \nnn{The mission consists} of two spacecraft, one carrying the instruments, and another one in formation flight at a distance of about 200\,m carrying the occulter to provide an artificial total solar eclipse. 
\nnn{This will ensure high-quality coronagraphic observations above the solar limb}

\nnn{\solmex\ integrates} two spectro-polarimetric coronagraphs for off-limb observations, one in the EUV and one in the IR, and three instruments for observations on the disk. The latter comprises one imaging polarimeter in the EUV for coronal studies, a spectro-polarimeter in the EUV to investigate the low corona, and an imaging spectro-polarimeter in the UV for chromospheric studies.

\nnn{SOHO and other existing missions have investigated the emission of the upper atmosphere in detail (not considering polarization), and as this will be the case also for missions planned for the near future. Therefore} it is timely that \solmex\ provides the final piece of the observational quest by measuring the magnetic field in the upper atmosphere through \emph{polarimetric} observations.

\end{abstract}

\section{Introduction}                                       \label{S:intro}

The Sun is our closest star and the only one for which we can resolve details on the surface, in its atmosphere, and in the astrosphere surrounding it. The direct impact the Sun has on the Earth and human life makes it the most important object in the sky. A total eclipse of the Sun is still one of the most breathtaking and captivating   wonders of nature. While in bygone ages it frightened whole societies, today scientists are intrigued by the beauty of the physics hidden behind the aesthetic pictures taken during these rare events.

During a total eclipse we see the dilute hot outer atmosphere of the Sun. Composed of million K hot plasma permeated by magnetic field, it presents us with a unique plasma physics laboratory, representing a parameter space not accessible on Earth. The magnetic field confines the coronal structures, drives the plasma in the upper solar atmosphere, provides the energy released in the form of thermal and kinetic energy, and controls the acceleration of particles.


Therefore, it is of pivotal importance to know the state of the magnetic field to reveal the nature of the governing processes in the outer solar atmosphere. So far measurements of the solar magnetic field are mostly restricted to the low layers of the solar atmosphere. Extrapolation techniques containing numerous assumptions are used to estimate the magnetic field in the transition region from the chromosphere to the corona and in the corona itself. Until now, few facilities provide measurement access to the magnetic field in the upper atmosphere, and each only represents a partial aspect and suffers from significant constraints and deficiencies. 

The \emph{measurement objective} of \solmex\ is to fill this gap of the observational puzzle and to present the first comprehensive set of magnetic field measurements of the upper solar atmosphere. 
\nnn{Such observations are the key to address the} main \emph{science objectives} of \solmex\ \nnn{which are summarized through the following questions:}

\begin{itemize}
\item[1.] What is the magnetic structure of the outer 
                solar atmosphere?

\item[2.] What is the nature of the changes of the magnetic 
                field over the solar cycle?

\item[3.] What drives large-scale coronal disruptions such 
                as flares and coronal mass ejections?

\item[4.] How do magnetic processes drive the dynamics and 
                heat the outer solar atmosphere?

\item[5.] How does the magnetic field couple the solar 
                atmosphere from the photosphere to the corona?
\end{itemize}

These scientific goals of \solmex\ are closely related to two of the four main questions of the \emph{Cosmic Vision} plan, namely ``\emph{how does the solar system work?}'' and ``\emph{what are the conditions for planet formation and the emergence of life?}''
Besides contributing to the understanding of the upper solar atmosphere as an astrophysical object in its own right, \solmex\ will provide essential information on the origin of ``space weather'', which affects life and the habitability on Earth as well as other solar system bodies. Unveiling the nature of the outer atmosphere of the Sun is the most important step in understanding the magnetised atmospheres of other cool stars. This is of particular interest for younger stars, because the magnetic field surrounding the star plays a decisive role in the formation of planets around them.


%
To reach \nnn{the objective to measure} the magnetic field in the upper atmosphere, \solmex\ will investigate the emission and its polarization from the extreme ultraviolet to the infrared. The magnetic field in the source region of the emission modifies the polarization through the Zeeman and Hanle effects. Measuring not only the spectral intensity, but also the polarization state, \solmex\ will provide the essential data to decipher the imprint of the magnetic field. Through inversion procedures and a comparison with numerical forward models \solmex\ gives access to the magnetic field in the upper solar atmosphere.

%
The key for the coronagraphic observations is to provide an artificial eclipse. To minimize stray light for the delicate polarimetric observations and to reach the required high spatial resolution, the occulter that blocks the light from the solar disk has to be far removed from the entrance aperture. \solmex\ will achieve this by mounting an occulting disk on a separate spacecraft that serves as an ``artificial Moon'' for the coronagraphs on the master spacecraft, which carries all science instruments. The spacecraft carry out a formation flight, separated by 200\,m..

We plan an orbit around the Lagrange point between Sun and Earth (L1), which provides an unobstructed view of the Sun at constant thermal conditions with the smallest possible impact on the formation flight. The anticipated mission lifetime spans a nominal phase of three years and an extended phase of three years. The nominal phase will cover the rise of solar activity, while the extended phase will monitor the solar activity maximum, assuming a launch in 2022.

%
\solmex\ comprises five instruments dedicated to study five prime target regions off-limb and on-disk. \vircor\ and \chrome\ are imaging spectro-polarimeters, \cusp\ and \susp\ are slit spectro-polarimeters. \eip\ provides imaging polarimetry data.
\begin{tabbing}
\emph{off-limb:}~~~~~ \=     \= \cusp\ ~~~~~~~  \= (Coronal UV spectro-polarimeter),\\
                      \>     \> \vircor\ \> (Visible light and IR coronagraph),\\
\emph{on-disk:}       \>     \> \eip\    \> (EUV imaging polarimeter),\\
                      \>     \> \susp\   \> (Scanning UV spectro-polarimeter),\\
                      \>     \> \chrome\ \> (Chromospheric magnetic explorer).
\end{tabbing}

\nnn{The \solmex\ mission is based on significant heritage from the \compass\ proposal \cite{Fineschi+al:2007} to a previous M-class call by ESA. The concept of \cusp, \vircor\ and \susp\ follows the \compass\ proposal, but \eip\ and especially \chrome\ provide new developments. While the basic idea for the mission remains similar, the two spacecraft for the formation flight are a completely new design and the mission profile has been developed again from scratch.
}

The scientific objectives are summarized in \sect{S:sci.objectives} \nnn{and} the basic measurement techniques are briefly presented in \sect{S:observables}. In \sect{S:B-diagnostics}  the five prime target structures and the relevant diagnostics to measure the magnetic field are described. 
The mission profile with two spacecraft in formation flight is presented in \sect{S:mission}. Section \ref{S:payload} describes the details of the five payload instruments and \sect{S:spacecraft} gives some details on the required spacecraft. Section \ref{S:conclusions} concludes the paper.

\section{Scientific objectives}  \label{S:sci.objectives}

One of the greatest challenges in space astrophysics in the immediate future will be the empirical investigation of the magnetic field vector in a variety of astrophysical plasmas, such as the solar corona, circumstellar envelopes, accreting systems, etc. In particular, we need to decipher the three-dimensional magnetic structure of the solar upper atmosphere. \nnn{This is because the magnetic field (a) is the source for the energy to heat the corona, (b) is governing the transformation of magnetic into thermal energy and (c) is driving and channelling the dynamics in the upper atmosphere.}

\begin{figure}
\includegraphics[width=\textwidth,bb=26 31 515 199,clip=true]{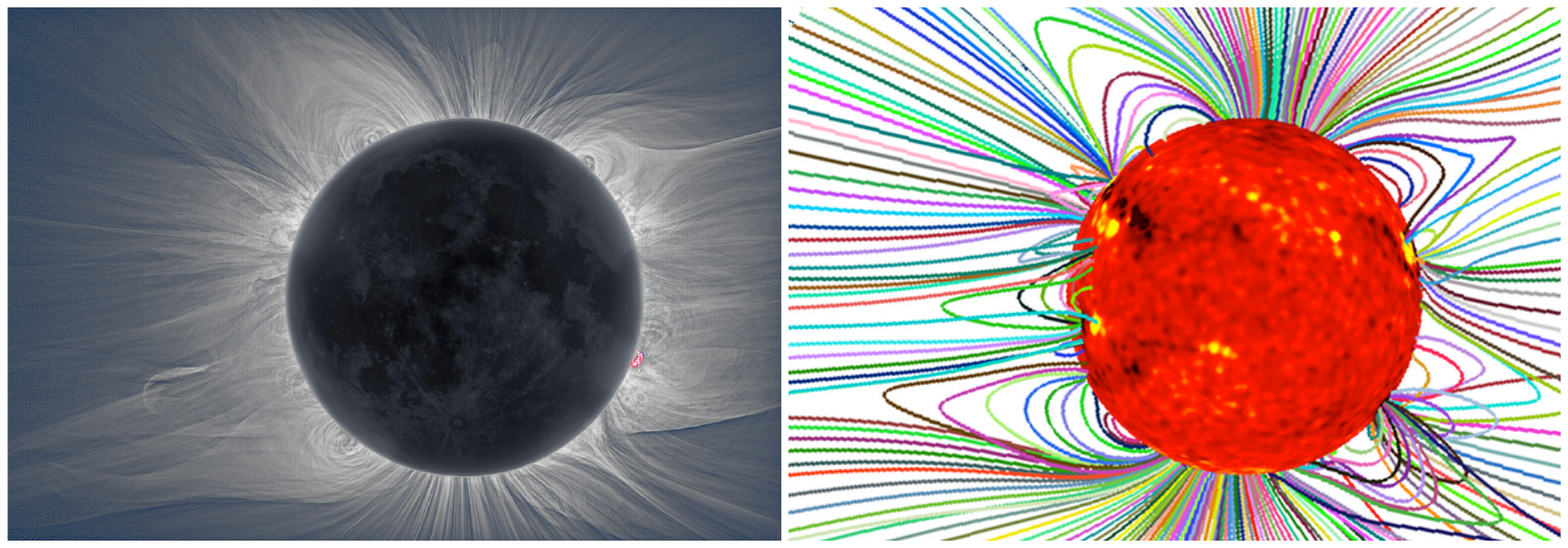}
\caption{\emph{Left panel:} Total solar eclipse of 11 Jul 2010 (courtesy Druckm\"uller, Dietzel, Habbal, Ru\v{s}in; http://www.zam.fme.vutbr.cz/$\sim$druck/eclipse/).
The image is contrast enhanced. 
%
%
\emph{Right panel:} Magnetic field lines as following from a global MHD model for the same moment as the eclipse observation on the left. On the solar surface a magnetogram of the photospheric field is shown (courtesy Linker, Miki\'c, Lionello, Riley, Titov; http://www.predsci.com/corona/jul10eclipse/jul10eclipse.html).
%
%
\label{F:intro}}
\end{figure}

The upper atmosphere of the Sun is highly structured, which is evident in contrast-enhanced eclipse photographs (cf.\ \fig{F:intro}, left panel) and from images in the extreme ultraviolet. In fact, this is due to the magnetic field, but our knowledge of the magnetic field in the upper solar atmosphere mainly depends on \emph{static} extrapolations from the solar surface (cf.\ \fig{F:intro}, right panel). While this technique is useful for a number of science questions, there is the acute need to actually \emph{measure} the magnetic field and its evolution in the \emph{dynamic} solar upper atmosphere.

Our current lack of measurements of the magnetic field in the upper chromosphere, transition region, and corona is a major obstacle to our understanding of the outer solar atmosphere. Consequently, the main scientific goal of the proposed mission is to provide the first comprehensive set of measurements of the magnetic field in the chromosphere, the transition region to the corona, and the corona itself through remote-sensing techniques over a broad range of spatial and temporal scales.
%


\nnn{In this section} we will limit the discussion of the scientific objectives to a summary in tabular form and a few examples.
Table\,\ref{T:sci} shows how the five major science questions are subdivided into more detailed questions and how these map to the measurements of the \solmex\ instruments (\sect{S:payload}) of the prime target regions (\sect{S:B-diagnostics}).

\nnn{The combined measurements on different temporal and spatial scales will investigate the magnetic coupling of various solar structures. This is central to achieve the scientific goals.}

\subsection{Science example 1: Magnetic structure of coronal loops}

\begin{figure}
{\includegraphics[width=\textwidth,bb=26 31 483 201,clip=true]{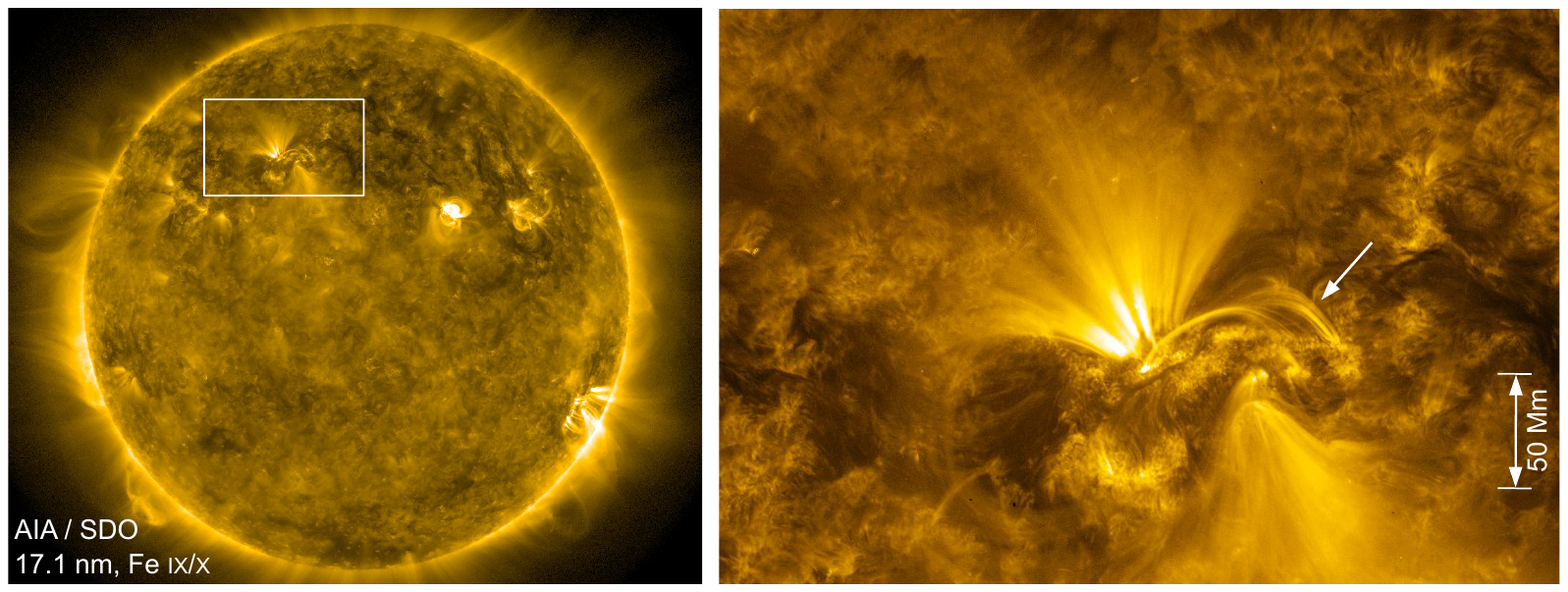}}
\caption{%
The corona at 10$^{\,6}$\,K seen in the extreme ultraviolet. The right panel shows an enlargement of the framed region in the left panel. \solmex\ will not only provide intensity images like these, but also data on the magnetic field direction, which is important to understand, e.g., the nature of coronal loops (arrow in right panel).
Image courtesy SDO/AIA team.
\label{F:sci-sdo}
}

\end{figure}

Imaging observations show loops  in great detail that reach high into the corona, up to altitudes of the order of 100\,Mm ($\approx$10\% of the solar radius) or more (\fig{F:sci-sdo}). It is generally accepted that these outline magnetic field lines, which are loaded with hot plasma as a consequence of the coronal heating process.  Furthermore, the magnetic structure of coronal loops is unknown, e.g., their twist and their magnetic connection to the lower atmosphere are open issues.

\solmex\ will measure not only the appearance and dynamical evolution of the loops in coronal extreme UV emission, but will also measure the strength and direction of the magnetic field along the loop and at its base where it is rooted to the chromosphere. Furthermore, coronagraphic observations with \solmex\ will provide eclipse-like images (see \fig{F:intro} for an eclipse image) with unprecedented quality very close to the limb, and this will enable us to study the evolution of coronal loops, e.g., above active regions, in hitherto unknown detail.

\subsection{Science example 2: Spicules and their coupling to the corona}

\begin{figure}
\centerline{\includegraphics[width=0.85\textwidth,bb=26 30 368 160,clip=true]{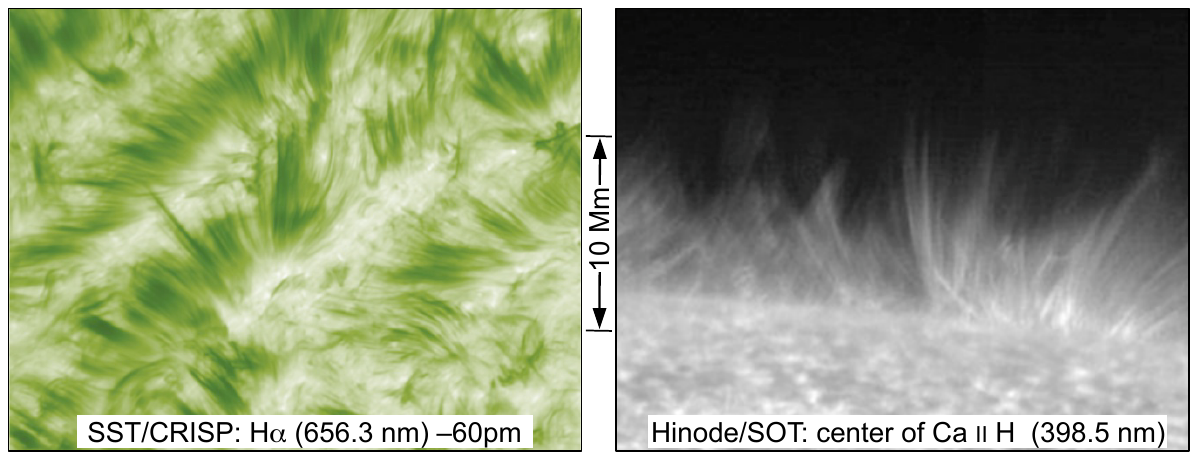}}
\caption{%
Spicules on the disk (dark, left) and at the limb (bright, right).
\chrome\ on \solmex\ will deliver similar images, but now including measurements of the magnetic field in these thin structures.
Left panel: courtesy of A. Pietarila, right panel from \cite{Judge+Carlsson:2010}.
\label{F:sci-spicule}
}

\end{figure}

The Sun is highly structured and very dynamic on smaller scales of the order of 1\,Mm in the chromosphere and in the transition region to the corona. A multitude of small-scale short-lived phenomena can be seen, such as spicules, fibrils, explosive events, etc. An example of spicules seen on the disk and above the limb is shown in \fig{F:sci-spicule}.  Spicules have been extensively studied with Hinode in the chromospheric emission of \ion{Ca}{2}. Indications of Alfv\'en waves that travel along the spicules have been found \citep{DePontieu+al:2007}. Recent observations provide evidence of high-velocity upflows in type-II spicules, which feed mass into the hot corona and thus are an integral part of the coronal heating problem \citep{McIntosh+DePontieu:2009:upflows,Peter:2010}.
However, all these conclusions are based on intensity images alone, and with ground-based telescopes only low-resolution measurements of the magnetic field in spicules could be performed \citep[][and references therein]{Centeno+al:2010}. At somewhat higher temperatures, there are small transition region loops found crossing the chromospheric network. These are puzzling because there appears to be no connection from these small network loops to the underlying photospheric magnetic field \citep{Feldman+al:2001}.

\solmex\ will provide key data to address these questions by inferring the magnetic field 
from the observed polarization in lines formed in the chromosphere and the transition region. Measurements of the magnetic field in \ion{Mg}{2}\,k (279.6\,nm), \Lya, and \ion{C}{4} (154.8\,nm) will yield magnetic information on spicules on the disk and above the limb, and on small transition region loops above the chromospheric network. Such observations will uncover the role of such features in heating the plasma. The coronagraphic instruments will provide the crucial connection to the structures in the corona.

\subsection{Science example 3: magnetic structure of a CME during an eruption}

\begin{figure}
\parbox[b]{0.55\textwidth}{\includegraphics[width=0.55\textwidth]{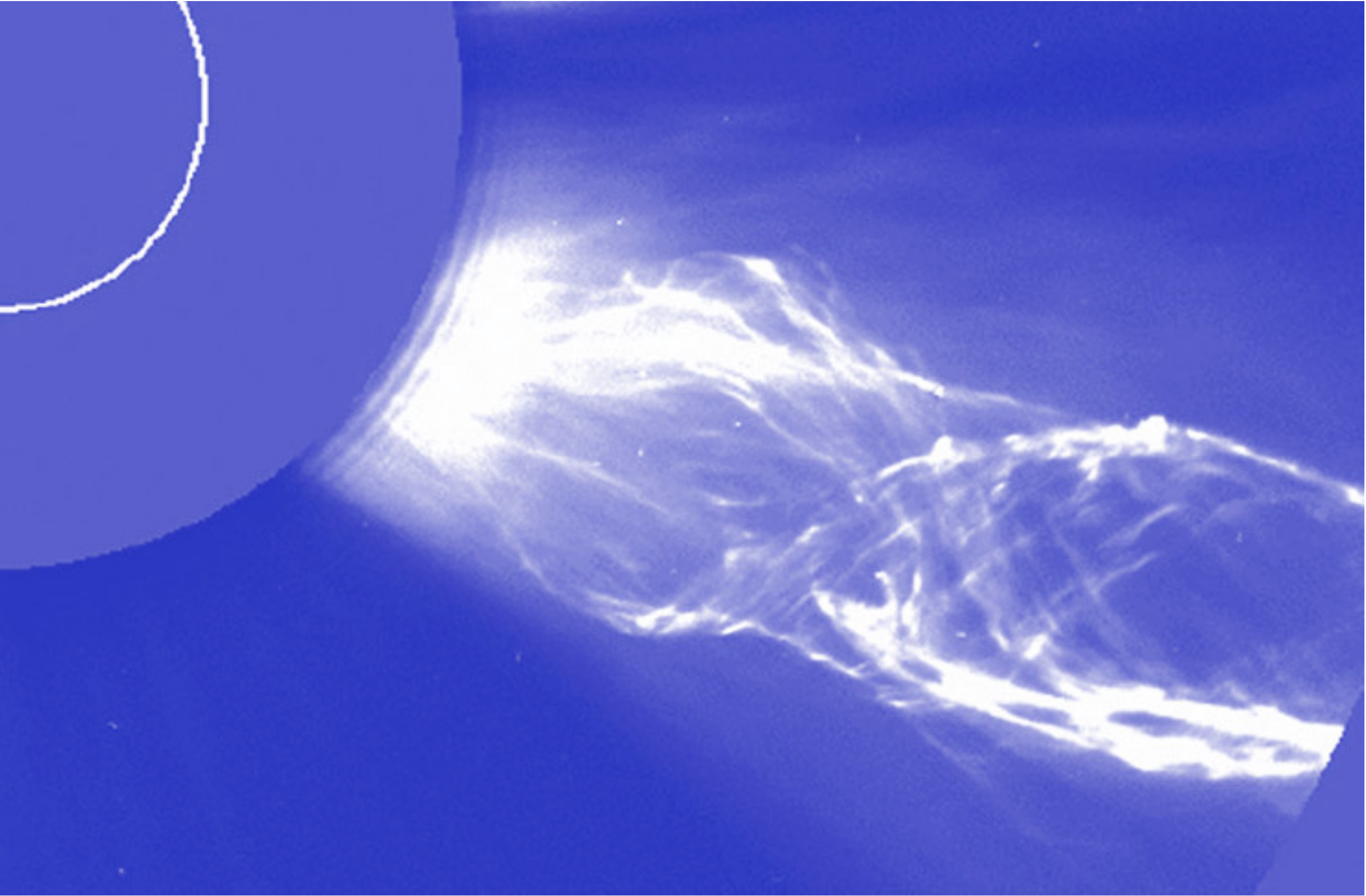}} \hfill
\parbox[b]{0.40\textwidth}{
\caption{%
Coronal mass ejection (CME) with a helical structure observed by Lasco\,C2 on SOHO. Through spectro-polarimetric observations \solmex\ will allow us to measure the magnetic field in these structures for the first time. (The thin white line outlines the solar limb).
\label{F:sci-cme}
}}
\end{figure}

Once a coronal mass ejection (CME) lifts off, it expands, and in the process some CMEs reveal their internal structure (\fig{F:sci-cme}). Our current information is based on intensity images, thus no information on the magnetic field is available. This is unfortunate because competing models for CME eruption make detailed predictions on their internal magnetic structure, e.g., breakout, flux rope catastrophe, or kink instability \citep{Forbes+al:2006}. As CMEs often lift off with speeds of the order of 1000\,km/s, they cross a field of view of one solar radius in 10 minutes, which sets minimum requirements for the cadence of observations.

Through high-cadence infrared coronagraphic spectro-polarimetry \solmex\ will provide outstanding data that will not only reveal the density structure of CMEs, but also return the magnetic field vector. This includes the phases before, during and after an eruption, as the spectro-polarimetric coronagraph can reach a cadence of down to 30\,s, well below the crossing time of a CME through the field-of-view. This will decide the debate of the different CME models and show the true physics of the CME eruption.

\begin{table*}[t]
\caption{Summary of science goals and their relation to the prime target structures and observational techniques and the \nnn{measurement requirements for the} instruments (\sects{S:B-diagnostics} and \ref{S:payload}).
           \label{T:sci}}%
{
}
\includegraphics[width=\textwidth,bb=14 58 581 786,clip=true]{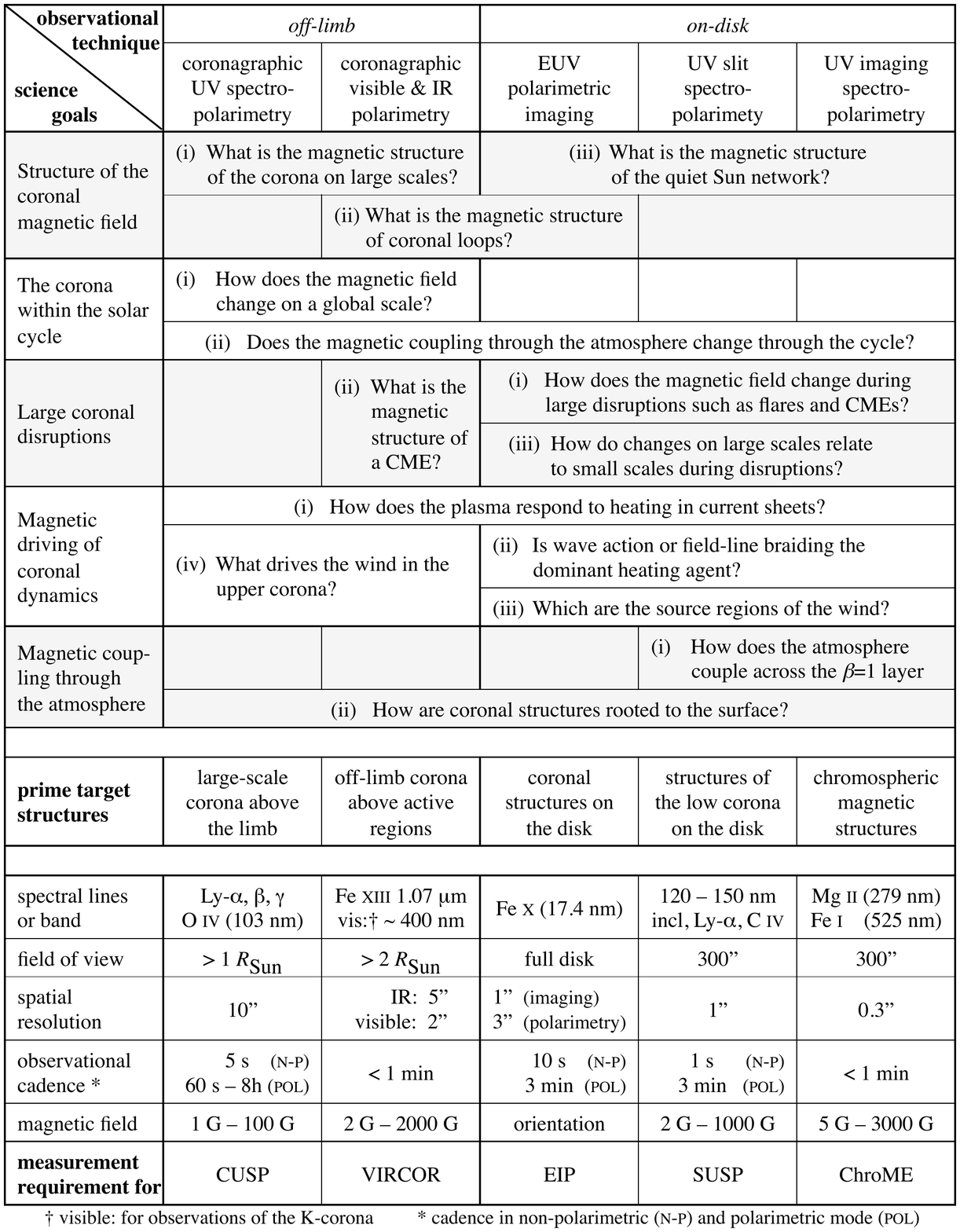}
\end{table*}

\section{Observable parameters and measurement techniques\label{S:observables}}

The only means of investigating the upper atmosphere of the Sun at distances of below a couple of solar radii above the surface is through remote sensing. We can detect the direction, flux, energy, and state of polarization of the incoming photons.

To observe the corona in the visible and infrared (IR), one has to block the light from the solar disk by an occulter (creating an artificial eclipse) because otherwise it would outshine the faint emission from the corona (see \sect{S:mission}). In contrast, observations in the ultraviolet (UV) and extreme UV (EUV) range also show the transition region and the corona in front of the solar disk, because the cool photosphere is relatively dark at these short wavelengths. Therefore, UV/EUV instruments and coronagraphs are complementary.

The emission of the solar coronal plasma in the EUV (and X-rays), observable only from space, has been studied extensively by, e.g., SOHO, TRACE, Hinode, and SDO. These missions provided high-resolution intensity images and spectra. They have revolutionised our understanding of the outer solar atmosphere, because they allowed quantitative measurements of the kinetic temperature, composition, density, and dynamics of the coronal plasma. However, the information on the coronal magnetic field is encoded in the polarization of the spectral lines, and was therefore inaccessible to previous space instruments. 

For the first time the \solmex\ instruments will measure the polarization of light from the chromosphere and the corona from space, on-disk as well as off-limb. Thus, they will provide a comprehensive set of measurements of the coronal magnetic field. Besides this unique capability, all instruments perform as well or better than previous and existing space-borne instruments of their category when operated in non-polarimetric mode. 
The details of these instruments are discussed in \sect{S:payload} and summarized in \tab{T:instr}

\paragraph{The Stokes vector,} $(I,Q,U,V)$, is used to describe the polarization state of the light \citep{Stix:2002}. Here $V$ is the intensity difference between right and left circular polarization, $Q$ is the intensity difference between linear polarization parallel and perpendicular to a given reference direction, and $U$ is the intensity difference between linear polarization at $+45^\circ$ and $-45^\circ$. $I$ is the total intensity at the wavelength under consideration.
\solmex\ will measure the Stokes vectors of near IR, optical and EUV spectral lines and use the Zeeman and Hanle effects to determine the vector magnetic field.

\paragraph{The Zeeman effect} is a manifestation of the interaction of the magnetic moment of the atom and the magnetic field. In the presence of a magnetic field, the spectral lines split into components with circular and linear polarization, depending on the angle between the line-of-sight and the magnetic field vector.   
The Zeeman effect is more prominent in circular polarization (longitudinal Zeeman effect), and particularly in the infrared (IR). 
It is a valuable tool for coronagraphic observations above the limb \citep{Harvey:1969}.
Circular polarization is measurable in some chromospheric and low coronal ultraviolet (UV) lines also when pointing to active regions on the solar disk, although with smaller amplitudes. Previous attempts to measure polarization in UV lines did not achieve the necessary signal-to-noise ratio (see \sect{S:TR}). The most recent attempt with the SUMI rocket (summer 2010) unfortunately did not provide polarimetric data. \solmex\ is designed to overcome the previous shortcomings.


\paragraph{The Hanle effect} is the magnetic modification of atomic level polarization (population imbalances and quantum coherences among the sublevels of degenerate atomic levels), which results from the absorption of anisotropic radiation. The (magnetically modified) atomic level polarization leads to linear (scattering) polarization of the emitted light. Measuring this polarization gives access to the magnetic field.
Depending on the scattering geometry, the modification consists of a reduction or enhancement of the linear polarization amplitude and in a rotation of the direction of linear polarization. In general, the Hanle effect in ultraviolet (UV) permitted lines (e.g., \Lya) is sensitive to the strength and orientation of the magnetic field vector. In coronal forbidden lines, the Hanle effect saturates and is sensitive to the field orientation only \citep{Charvin:1965}. This is one of the reasons why the Hanle effect diagnostics in permitted UV and forbidden IR lines are complementary and their combination gives access to a large range of magnetic field strengths.

\section{Prime target structures and magnetic field diagnostics with \solmex} 
\label{S:B-diagnostics}

\solmex\ will use five different observational techniques to derive the magnitude and direction of the magnetic field. 
Each of these techniques is best suited for one prime target structure as detailed below. \tab{T:sci} shows how the observational techniques relate to the prime targets of the \solmex\ instruments and the science questions. 
While all instruments for \solmex\ will also provide traditional diagnostics well known from previous space missions, e.g., density or velocity diagnostics, we will concentrate in the following on the new diagnostic tools to measure the magnetic field in the prime target structures.

\subsection{Large-scale corona above the limb} \label{S:large.scale.corona}

\begin{figure}[t]
\parbox[b]{0.60\textwidth}{%
    \includegraphics[width=0.60\textwidth,bb=26 31 254 159,clip=true]{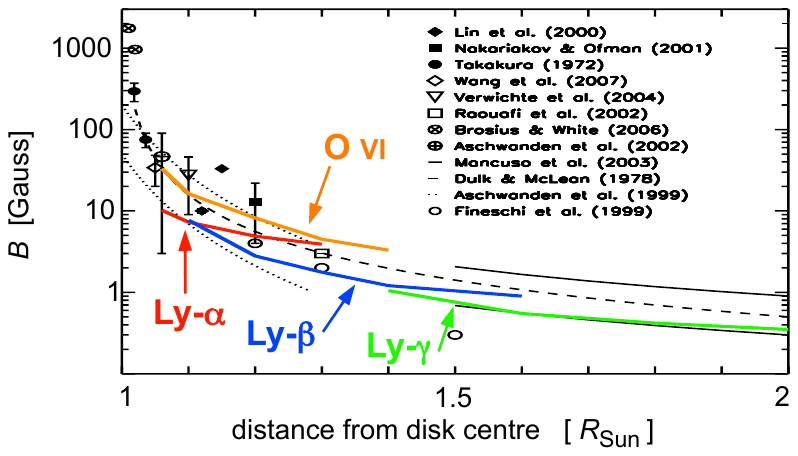}} \hfill
\parbox[b]{0.35\textwidth}{\caption{%
Average magnetic field in the corona above the limb as derived from different sources. Over-plotted in colour are minimum field strength sensitivities of the emission lines to be observed by \cusp\ on \solmex.
\label{F:polarization-Ly}
}}
\end{figure}

The large-scale corona above the limb, e.g. at the boundaries between coronal holes and quiet Sun or in coronal streamers, can be expected to show only weak magnetic fields of 1\,G to 100\,G (\fig{F:polarization-Ly}). 
This requires the use of the Hanle effect in 90$^\circ$ scattering geometry, which also has the  advantage that it is sensitive to magnetic fields that are randomly oriented within the observational spatio-temporal resolution element (contrary to the Zeeman effect polarization). The best-suited spectral lines are found in the UV-range, the lines of the Lyman series of H and \ion{O}{6} at 103.2\,nm \citep[e.g.][]{Fineschi:2001}. Above the limb these show high linear polarization (up to 25\%), which is modified by the magnetic field \citep{Bommier+Sahal-Brechot:1982}.
Using a roll of the SOHO spacecraft, SUMER was used to estimate the linear polarization of the O VI 103.2 nm line via intensity modulation as a function of the roll angle [Raouafi et al. 1999]. 3G were found at 0.3 R above the limb, which proved the measurement concept [11].

Using a roll of the SOHO spacecraft, SUMER was used to estimate the linear polarization of the \ion{O}{6} 103.2\,nm line via intensity modulation as a function of the roll angle\citep{Raouafi+al:1999}. The derived polarisation corresponded to 3\,G at 0.3\,$R_\odot$ above the limb, which proved the measurement concept \citep{Raouafi+al:2002}. 

The full potential of this method can only be reached by a dedicated UV spectro-polarimeter for off-limb observations that occults the UV light from the solar disk. Only these coronagraphic observations allow a sufficiently low level of instrumental scattered light.
This will be done by \cusp\ on \solmex, reaching signal-to-noise ratios of 100 in polarimetric measurements of UV lines, thus providing sufficient sensitivity for the expected magnetic field strengths (\fig{F:polarization-Ly}).



\subsection{Off-limb corona above active regions} 

This target is best measured by an infrared (IR) coronagraph with an imaging spectro-polarimeter.  At the high magnetic field strengths of active regions, the IR lines show a large splitting of the Zeeman components. This allows us to deduce the direction and the strength of the magnetic field, if the linear polarization signal produced by scattering processes is also taken into account \citep{Casini+Judge:1999}. Because the corona above an active region (when located near the limb) is bright above the limb, temporal changes on the time scale of minutes can be detected making it possible to study Alfv\'en waves, i.e., distortions of the magnetic field.

These measurements have been performed on the ground, e.g., with the COMP \citep{Tomczyk+al:2007}. They showed some 30\,G at 0.05\,$R_\odot$ above the limb 
\citep[][]{Lin+al:2000} 
and indicate that the polarimetric accuracy has to be about $10^{-4}$ (circular and linear polarization compared to the total intensity) to obtain reliable values for the magnetic field. Ground-based observations suffer from a high level of stray light in the Earth's atmosphere, limiting the signal-to-noise ratio, and thus the polarimetric accuracy. This prevents observations higher than ${\approx}0.15\,R_\odot$  above the limb \citep[][]{Lin+al:2004}, and underlines the pressing need for space-based observations.

\begin{figure*}
\centerline{\includegraphics[width=90mm,bb=26 31 544 292,clip=true]{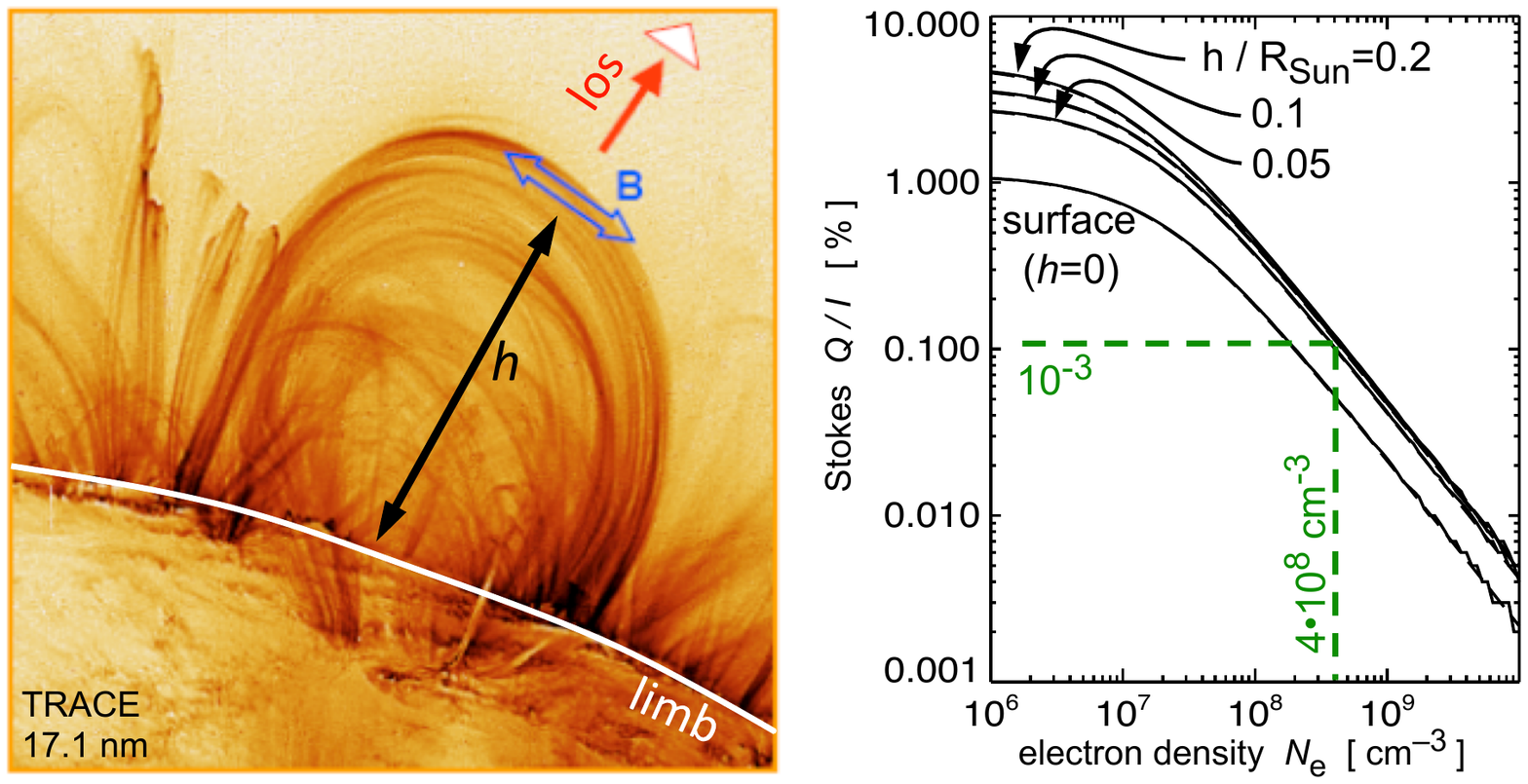}}
\caption{%
Predicted linear polarization signal $Q/I$ of the coronal \ion{Fe}{10} line at 17.4\,nm for on-disk observations for different heights $h$ above the surface.~
With densities of some 10$^{\,8}$\,cm$^{\,-3}$, a polarimetric signal $Q/I$ of well above 10$\,^{-3}$ can be expected for typical coronal loops, which will be detectable with \eip\ on \solmex. \citep[See][]{Manso-Sainz+TB:2009}.
The left panel shows a TRACE coronal image similar to what we expect to see in the intensity channel of \eip.
\label{F:FeX-polarization}
}
\end{figure*}

\subsection{Coronal structures on the disk}  \label{S:ondisk.corona}

On-disk coronal structures, including those located close to the solar limb, are best observed in permitted extreme UV (EUV) lines produced by the $10^6$\,K coronal plasma. This was done with great success by EIT/SOHO, TRACE, or most recently AIA/SDO by high-resolution imaging. But they provide information on the intensity only.

It has been shown that permitted EUV 
lines of \ion{Fe}{10} (formed around $10^6$\,K) should display linear polarization signals 
that can be used to infer the direction of the magnetic field in coronal loops and arcades 
\citep{Manso-Sainz+TB:2009}. The ground level of \ion{Fe}{10} is strongly   
polarized by anisotropic radiation pumping in a forbidden-line transition at visible wavelengths (where the solar disk is very bright). A significant fraction of this atomic level polarization can be transferred to the upper level of the permitted \ion{Fe}{10} line at 17.4\,nm by collisional excitation. 
The spontaneous emission from the thus polarized upper level then generates linearly polarized emission in the EUV line of \ion{Fe}{10} at 17.4\,nm (i.e., a wavelength at which the underlying solar disk is practically dark). 
For magnetic fields above 0.01\,G the lower level of the EUV line is in the Hanle saturation regime. Therefore, we can derive the direction of the magnetic field from the linear polarization signal (but not its strength).

A first estimate of the ratio of radiative to collisional excitations in the \ion{Fe}{10} 17.4\,nm line suggests that the linear polarization caused by scattering in the EUV line itself is less significant than that caused by the polarizing mechanism explained above.

For typical coronal structures with densities up to $10^9$\,cm$^{-3}$ the linear polarization signal of the \ion{Fe}{10} line at 17.4\,nm 
is expected to be a fraction of a percent (cf.\ \fig{F:FeX-polarization}), which is detectable with the imaging polarimeter \eip\ proposed here for SolmeX.

\begin{figure*}[t]
\centerline{\includegraphics[width=95mm,bb=26 31 326 139,clip=true]{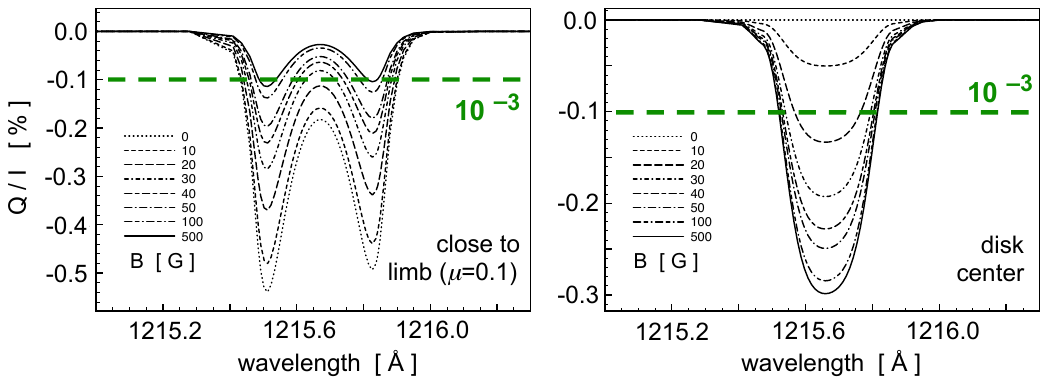}}
\caption{%
Predicted linear polarization signal of \Lya\ for on-disk observations.
\newline
For different strengths of a horizontal magnetic field the expected $Q/I$ profile is plotted for a close-to-the-limb observation (left) and for disk centre (right). The dashed line shows the polarimetric accuracy of \susp\ for \solmex. Figures from \cite{Trujillo-Bueno+al:2011}. See also \cite{Trujillo-Bueno:2011}.
\vspace{-2.0ex}
\label{F:Lya-polarization}
}
\end{figure*}

\subsection{Structures of the transition region and low corona on the disk} \label{S:TR}

This target is best observed with a UV spectro-polarimeter. This instrument has to combine good spectral coverage and spectral purity with the ability to image an appropriate region. For on-disk observations two suitable lines are \Lya\ at 121.5\,nm and \ion{C}{4} at 154.8\,nm. Both are formed in the transition region from the chromosphere to the corona at different heights, with individual structures reaching well into the corona up to some 30\,Mm.

Scattering processes in the solar transition region plasma are expected to produce linear polarization in \Lya, whose modification via the Hanle effect reveals the magnetic field \citep[see][]{Trujillo-Bueno+al:2011}. While close to the limb depolarization is expected, in the forward-scattering geometry of a disk centre observation the Hanle effect of inclined magnetic fields creates linear polarization \citep{Trujillo-Bueno+al:2002}. In both cases we can expect linear polarization of the order of 0.5\% for \Lya\ 
(see \fig{F:Lya-polarization}). Measuring the Hanle effect in \Lya\ is the main objective of the recently-proposed CLASP sounding rocket experiment \citep{Kobayashi+al:2010}.
%

\begin{figure*}[t]
\includegraphics[width=\textwidth]{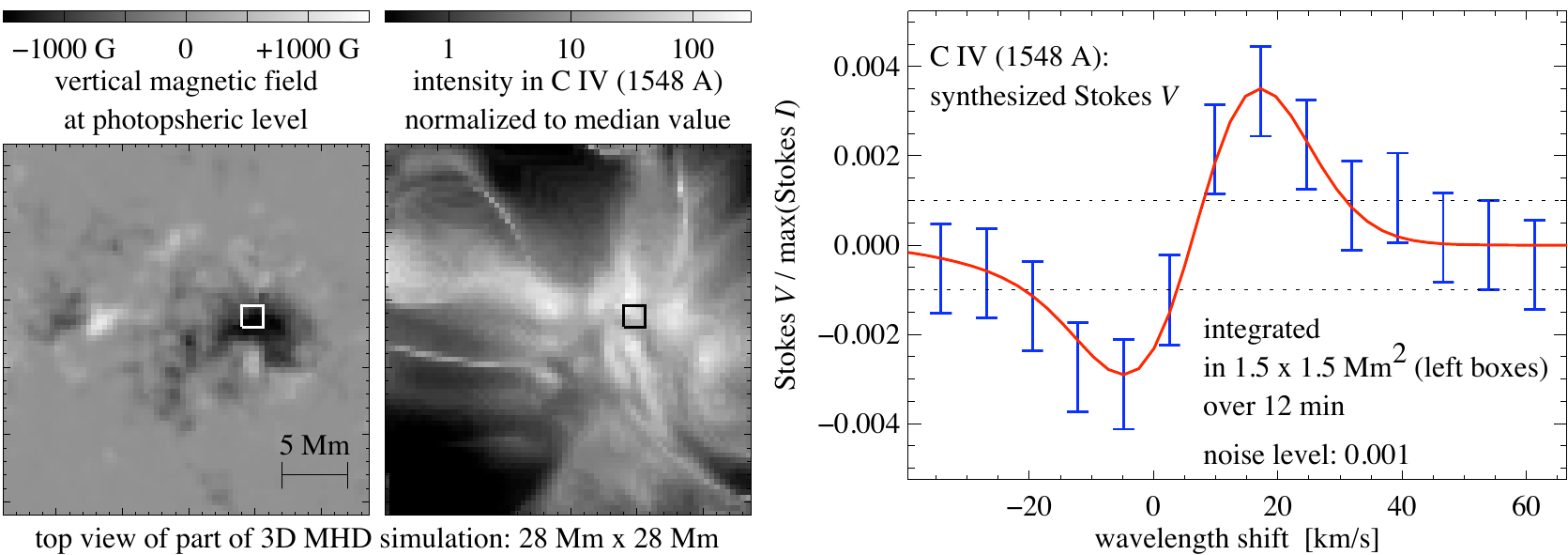}
\caption{%
Circular polarization signal in Stokes $V$ for  \ion{C}{4} (1548\,\AA) synthesized from a 3D MHD coronal model.
The left panels show the vertical magnetic field in the photosphere and the synthesized emission in part of a 3D MHD coronal model, looking at the computational domain from straight above. This part shows the surroundings of a pore. The synthesized Stokes $V$ profile shown to the right is acquired by averaging in time over 12 minutes and in space over 1.5\,Mm$\times$1.5\,Mm corresponding to {2\arcsec$\times$2\arcsec} as outlined by the small boxes in the left panels. The red curve shows the synthesized signal without noise. The bars represent the signal as it would be sampled with a noise of 10$^{-3}$ and a spectral resolution of about 40\,m{\AA}, which is comparable to the SUMER instrument. Data based on \cite{Peter+al:2006}.
\label{F:stokes.civ}
}
\end{figure*}

Concerning the \ion{C}{4} line at 154.8 nm, in regions with sufficiently strong magnetic field this line shows a measurable signature in circular polarization through the Zeeman effect. The level of polarization is also about a fraction of a percent. This can be expected in active regions, but also in smaller structures such as pores (see \fig{F:stokes.civ}) and possibly even in patches of strong network magnetic field in the quiet Sun. Previous measurements with {\sc uvsp/smm} a polarimetric accuracy just below 1\% \citep{Henze+al:1982,Hagyard+al:1983} gave no conclusive results except in sunspot umbrae \citep{Lites:2001}.
Based on the new more sensitive \solmex\ observations and with the help of models for the emission from the corona \citep[e.g.][]{Peter+al:2004}, the structure of the magnetic field can be inverted.

\subsection{Chromospheric magnetic structures} 

These structures can be revealed by spectro-polarimetric imaging in strong chromospheric lines. Among the most promising spectral lines for these diagnostics are the \ion{Mg}{2}\,h and k lines near 280 nm. Both lines are Zeeman-sensitive (effective Land\'e factors $g_{\rm{eff}}{=}1.17$ and 1.33). 
In addition, the \ion{Mg}{2}\,k line is expected to show strong linear polarization signals due to scattering processes, which through the Hanle effect are sensitive to the magnetic field of the upper solar chromosphere \citep[see][]{Trujillo-Bueno:2011}.
For \ion{Mg}{2}\,k one would measure the full Stokes vector, i.e., linear and circular polarization are recorded. This allows one to derive the full vector of the magnetic field (direction and strength). The strong absorption owing to the ozone band in the Earth's atmosphere requires this line to be studied in space outside the geocorona.

\begin{figure}[t]
\parbox[b]{0.60\textwidth}{\includegraphics[width=0.60\textwidth]{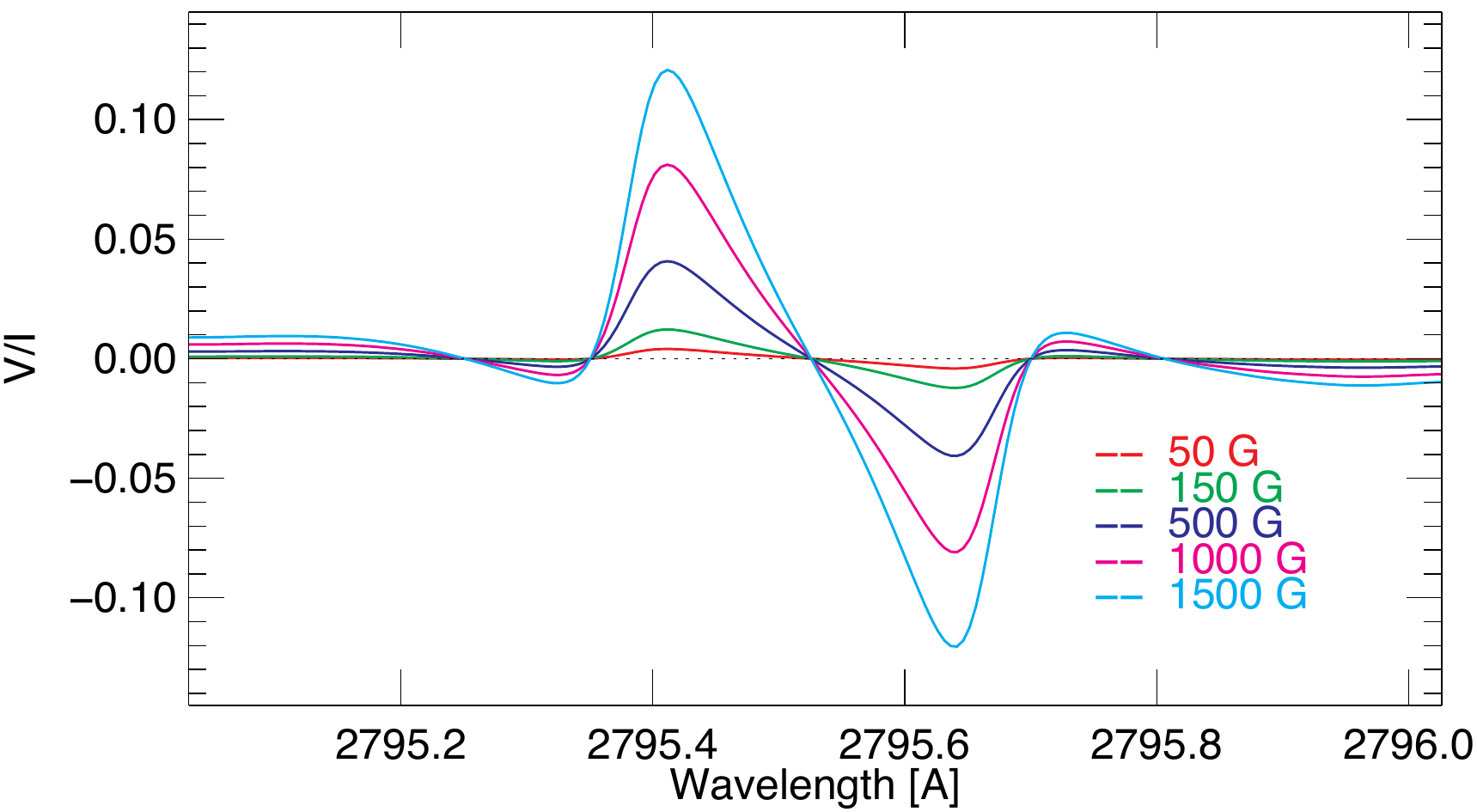}} \hfill
\parbox[b]{0.35\textwidth}{\caption{%
Circular polarization signal expected for the \ion{Mg}{2}\,k line at 279.6\,nm.
Only the very centre of the line profile is shown. The colored curves represent the Stokes $V/I$ signal for different magnetic fields along the line-of-sight. The curves are already degraded to show a synthetic profile similar to what can be expected from \chrome\ on \solmex.
\label{F:pol.chromo}
}}
\end{figure}

Based on 1D atmospheric models one can estimate the polarization signal from the \ion{Mg}{2} lines. As an example, \fig{F:pol.chromo} shows the circular polarization produced by the Zeeman effect assuming a magnetic field parallel to the line-of-sight (close to disk centre). 
For the typical magnetic fields of the upper chromosphere of active regions the expected circular polarization of \ion{Mg}{2}\,k is of the order of 1\%. In the quiet Sun chromosphere the linear polarization can be expected to be a few percent, which is sensitive to the weaker fields there via the Hanle effect \citep[][]{Trujillo-Bueno:2011}.

Because the chromospheric magnetic structures are rooted in the photosphere, co-temporal and co-spatial measurements of the photospheric magnetic field vector are required. This true for the comparison with the chromospheric channel, but also for the interpretation of \susp\ and \eip\ data. Such measurements can be achieved by recording the full Stokes vector, e.g., in the \ion{Fe}{1} line at 525\,nm.

\section{Mission profile: Two spacecraft in formation flight\label{S:mission}}

The measurement objectives require coronagraphic observations at \emph{low scattered light} level in the infrared, visible, and UV at \emph{high resolution} and \emph{close to the limb}. These three requirements, which are essential for the ambitious spectro-polarimetric observations in the solar corona above the limb, can only be achieved with an external occulter removed far from the entrance aperture of the instruments, because only this produces a true artificial eclipse.

\emph{This results in a mission profile with two spacecraft}, where one carries the instruments, and the other is the ``artificial Moon'', i.e. the occulter. 
\nnn{A similar profile is currently under investigation for the Proba-3 technology mission with its coronagraph {\sc{Aspiics}} \cite[e.g.][]{Vives+al:2006,Lamy+al:2008} and was employed for the proposal to the previous ESA M-class call DynaMICCS \citep{Turck-Chieze:2009}. However, the major science objectives of these missions are quite  different. As a technology mission Proba-3 contains only a limited set of instruments, and DynaMICCS was proposed to aim mainly at processes in the solar interior. 
\solmex\ as well as it predecessor \compass\ \citep{Fineschi+al:2007} are focussing on the processes in the upper solar atmosphere, most noticeably on the role of the magnetic field investigated through measurements of the polarization from the EUV to the IR.}

\emph{The scattered light level} we aim for with \solmex\ is about four orders of magnitude lower than for the leading ground-based coronagraph at Mauna Loa, Hawaii. The stray-light level of the Hawaii coronagraph of ${\approx}5{\cdot}10^{-6}$ is a factor of 5 better than what is envisaged for the high-resolution ground-based solar 4\,m telescope ATST in its coronagraphic mode. Unlike the ATST, which will enjoy its lowest stray-light level for only a short fraction of the time, \solmex\ will achieve its astonishingly low stray-light level of $10^{-10}$ virtually all the time. \nnn{This is possible through the combination of the external occulter, the refractive optics and the internal occulter and baffling.} Thus \solmex\ will provide a data quality surpassing that of planned ground-based facilities. \nnn{This is essential to observe not only the brightest coronal structures close to the limb, but also, e.g., faint large loops outside of active regions with sufficient signal-to noise for the polarimetric observations}.

\begin{figure}
\parbox[b]{0.75\textwidth}{%
  \includegraphics[width=0.75\textwidth,bb=26 31 351 253,clip=true]{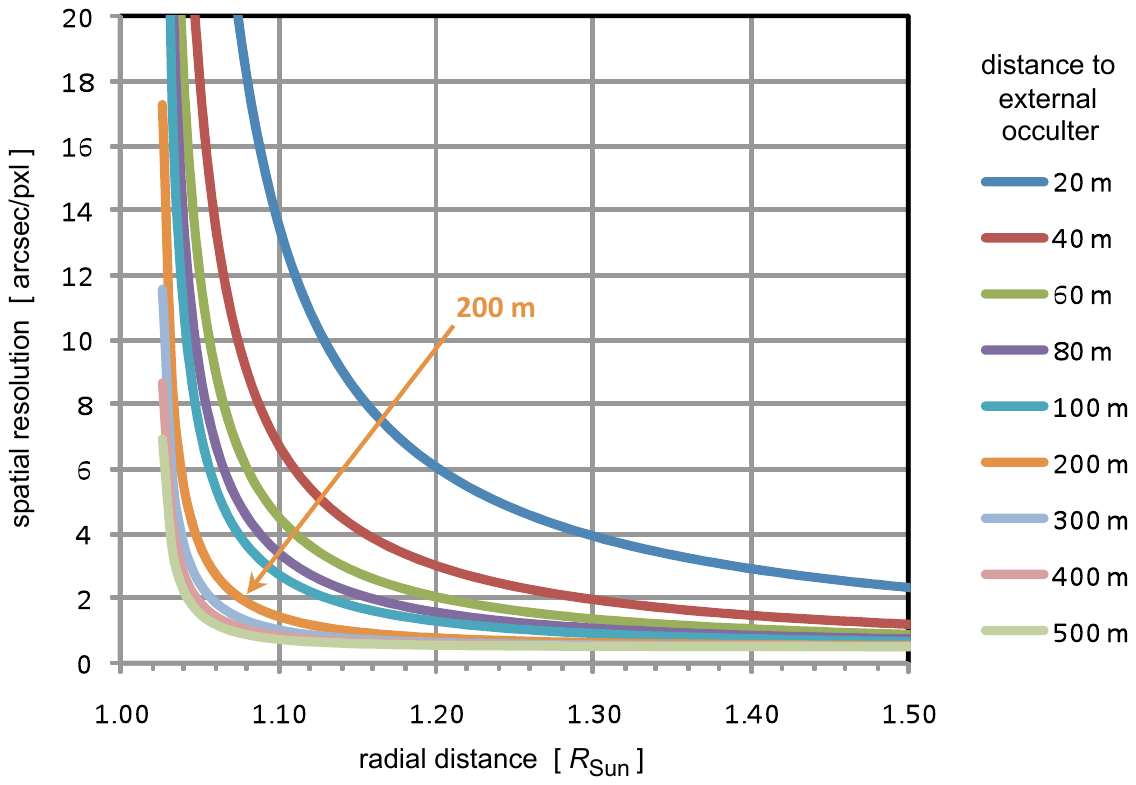}} \hfill
\parbox[b]{0.20\textwidth}{\caption{%
Spatial resolution that is possible for coronagraphic observations as a function of radial distance from the disk centre for different distances to the external occulter.
An occulter distance of 200\,m ensures a resolution of 2'' down to 0.07 solar radii or 50\,Mm above the limb.
(Courtesy of J. Davila).
\label{F:occulter}
}}
\end{figure}

\emph{High-resolution coronagraphic observations} are possible only for large occulter distances, because only then do the diffraction patterns induced at the occulter become sharp. This is illustrated in \fig{F:occulter}, which  shows the spatial resolution (as a function of distance from disk centre) for various occulter distances. Very close to the limb the resolution is poor because of the reduced effective aperture close to the occulter.

\emph{Observations close to the limb} are therefore difficult, and only possible at large occulter distances -- graphically speaking, the shadow has to have a distinct edge. These coronagraphic observations close to the limb are essential for understanding the magnetic coupling through the atmosphere and for achieving a connection to the on-disk observations, especially in the chromosphere.

For high-resolution observations (resolution element of 2'') of coronal loops that reach some 10\% of the solar radius into the corona, one has to observe at least as close as 50\,Mm to the limb. For an occulter removed about 200\,m from the entrance aperture this can be achieved (cf.\ \fig{F:occulter}), which also ensures the low stray-light level anticipated for \solmex.

Because an occulter removed 200\,m from the spacecraft cannot be kept sufficiently stable by a boom, the occulter and the telescopes of the coronagraphs need to be placed separately on two spacecraft.

\section{Instruments to measure the magnetic field in the upper solar atmosphere}  \label{S:payload}

\begin{table*}
\caption{Summary of the instruments discussed in \sects{S:instr.cusp} to \ref{S:instr.chrome}.
           \label{T:instr}}%
\includegraphics[width=\textwidth,bb=26 31 568 717,clip=true]{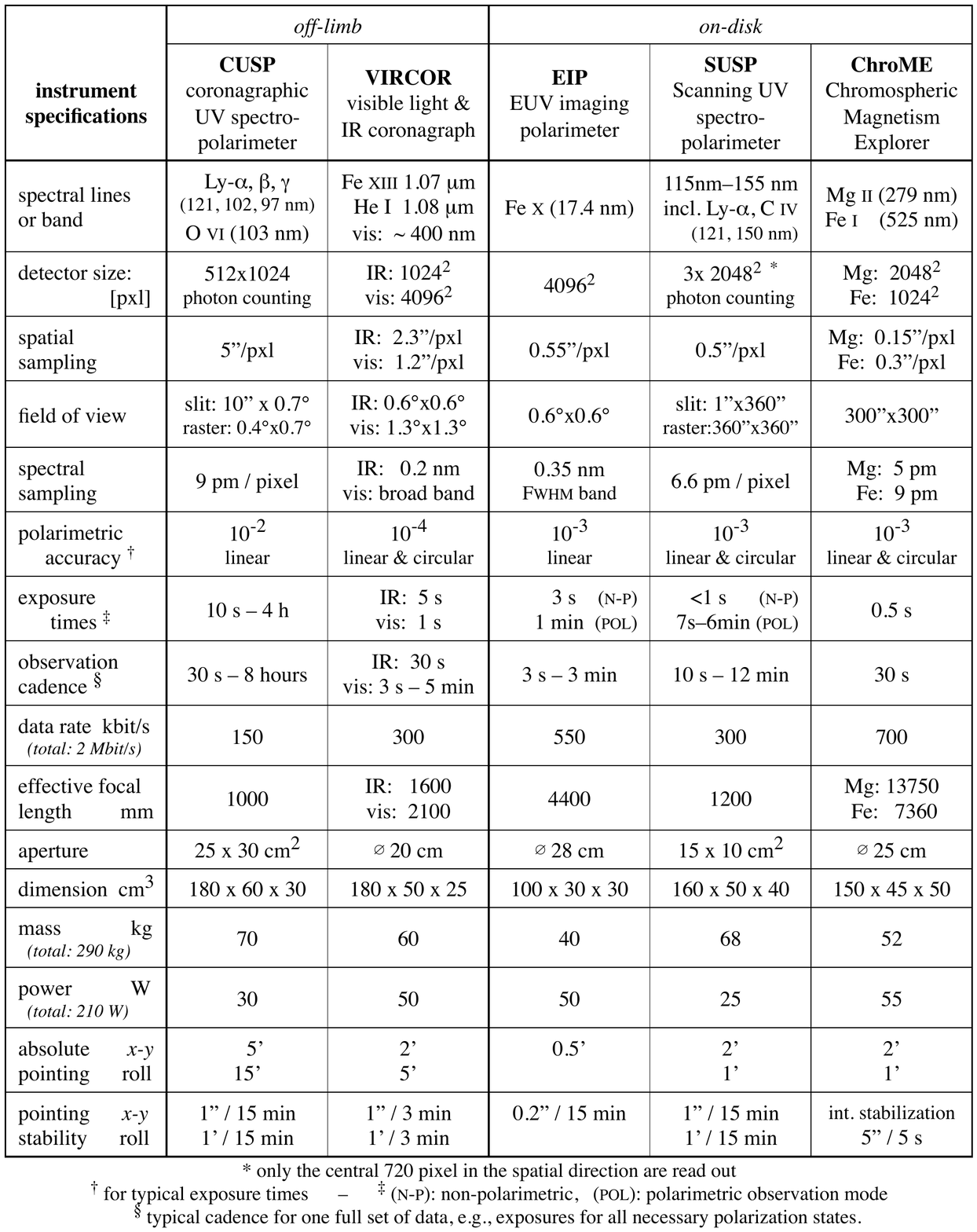}
\end{table*}

The suite of instruments for \solmex\ consists of five instruments. The two coronagraphs for off-limb observations use the 200\,m distant occulting disk on the separate occulter spacecraft to eclipse the solar disk. The three on-disk instruments investigate the chromosphere, transition region and corona in front of the solar disk in the UV and extreme UV light.

\begin{figure}
\centerline{\includegraphics[width=0.8\textwidth,bb=26 31 506 312,clip=true]{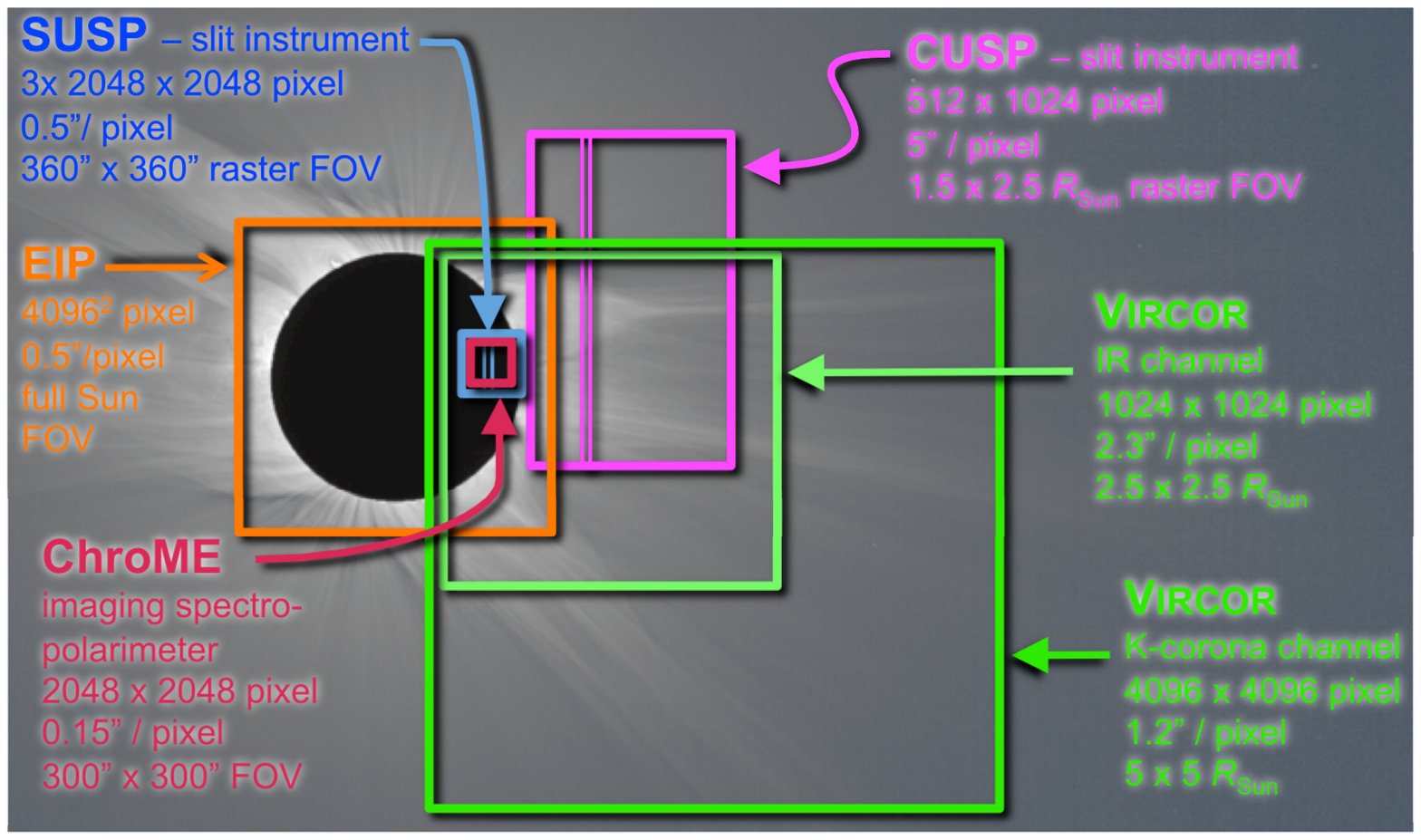}}
\caption{Fields of view of the \solmex\ instruments.
For the slit instruments \susp\ and \cusp\ a typical raster area is shown.
\susp\ can point anywhere on the disk and up to 1\,$R_\odot$ above the limb, \chrome\ anywhere on the disk. \cusp\ and \vircor\ can point to different parts of the corona through a roll of the spacecraft.\label{F:instr-fov}
}
\end{figure}

An overview of the technical details of the instruments can be found in \tab{T:instr}.
The field of view (FOV) of the \solmex\ instruments is depicted in \fig{F:instr-fov}. While the on-disk instruments either see the full disk or can point anywhere on the disk, the off-limb instruments need a simultaneous roll of the \solmex\ spacecraft to observe different parts of the corona.

\subsection{Coronal UV Spectro-Polarimeter (\cusp)}  \label{S:instr.cusp}


\cusp\ will measure the linear polarization of the solar UV radiation in the range of 95\,nm to 125\,nm in two channels, using an all-reflective polarization unit. The linear polarization signal is recorded in Ly-$\alpha$, $\beta$, and $\gamma$, 
and the \ion{O}{6} doublet at 103\,nm. This provides access to the magnetic field through the Hanle effect.

\begin{figure}
\centerline{\includegraphics[width=\textwidth,bb=26 31 538 205,clip=true]{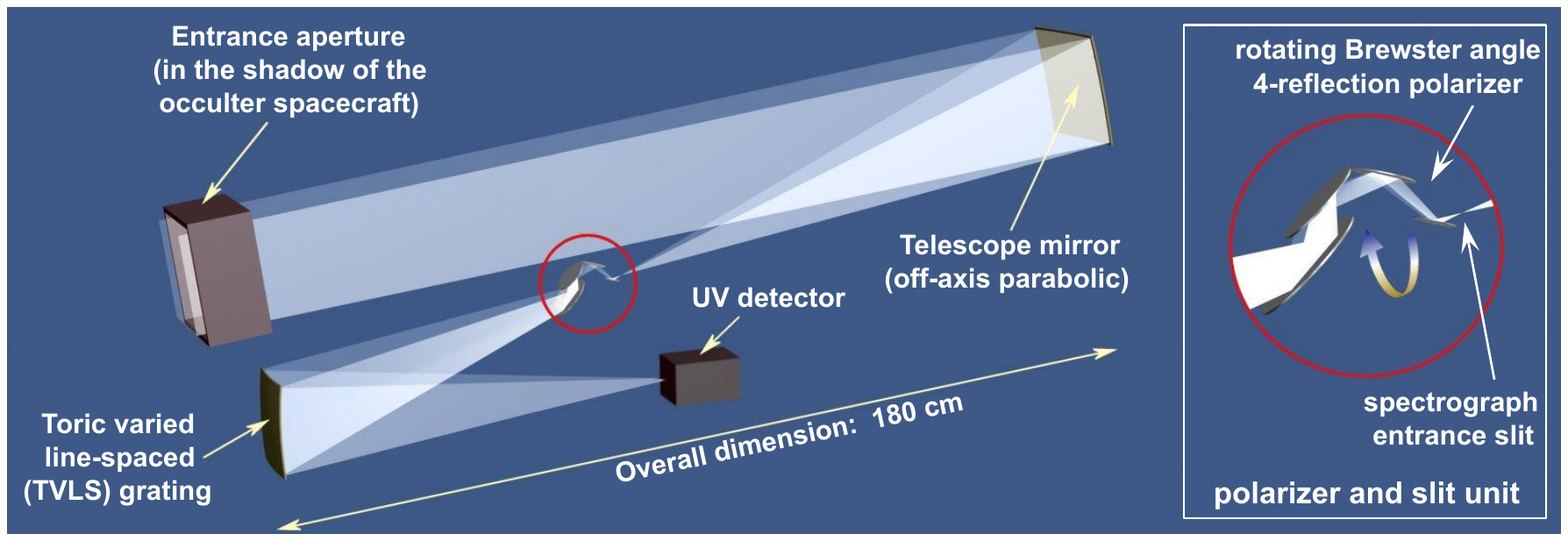}}
\caption{Schematic opto-mechanical layout of the Coronal UV Spectro-Polarimeter (CUSP).
           \label{F:instr-cusp}}%
\end{figure}


\cusp\ is an all-reflecting externally occulted coronagraph with the 200\,m long occultation baseline provided by the formation-flying occulter spacecraft. It uses the linear part of the occulter (\sect{S:spacecraft}) to minimize the effects of diffraction patterns.
Light from the solar corona enters the square entrance aperture and is imaged by a parabolic telescope mirror on the entrance slit of the spectrometer (\fig{F:instr-cusp}). Inside the spectrometer, the polarization is modulated by a rotating linear polarization analyser that consists of four reflecting plates at the Brewster angle \citep{Romoli+al:1994}, see inset of \fig{F:instr-cusp}. An all-reflective design is necessary because transmission optics such as those used for \susp\ (\sect{S:instr.susp}) work only above 105\,nm. This polarimeter configuration maintains the output beam coaxial with the input beam and can be retracted from the light beam. A toroidal varied line-spaced (TVLS) grating disperses the coronal radiation on the detector. 

The baseline detector is an Intensified CCD camera (ICCD) that is based on the suborbital experiment {\sc herschel}. A phosphorous screen is optically coupled to an image sensor, generally a fast-scan frame transfer CCD.
%
%
Coupling is achieved with a lens or a fiber optic taper which also provides the scaling required to match the CCD format to the phosphorous screen.


The sampling is 5"/pixel in the spatial and 9\,pm/pixel in the spectral direction. The spatial resolution is sufficient to investigate large-scale structures in the off-limb corona, such as streamers, or the boundary from the quiet Sun to coronal holes. The spectral resolution allows us to resolve the broad spectral profiles of coronal lines in detail. The 10" wide slit covers 0.7$^\circ$, corresponding to $2.5\,R_\odot$. The slit can be positioned to perform a raster scan by tilting the telescope mirror. A typical map will have a FOV of $0.4^\circ\times0.7^\circ$. 

\begin{figure}
\parbox[b]{0.55\textwidth}{%
   \includegraphics[width=0.55\textwidth,bb=26 31 254 161,clip=true]{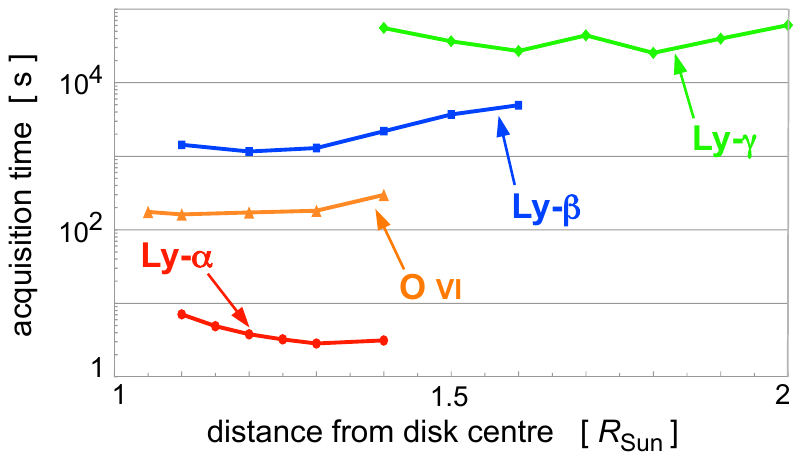}} \hfill
\parbox[b]{0.40\textwidth}{\caption{%
Typical acquisition time for \cusp\ for the prime target lines. 
This includes three exposures in three polarization states until a signal-to-noise ratio of 100 is reached.
A spatial binning of 5" was assumed for Ly-$\alpha$ and \ion{O}{6}, 10" for Ly-$\beta$, and 25" up to 100" for Ly-$\gamma$.
\label{F:instr-cusp-exp}
}}
\end{figure}

\cusp\ will record the best-suited UV lines for magnetic field diagnostics: the H-Lyman series and \ion{O}{6} (\sect{S:large.scale.corona}). Because the expected degree of polarization is 1\% to 25\%, a signal-to-noise ratio of at least 100 has to be achieved. The required exposure times (for different spatial binning) are shown in \fig{F:instr-cusp-exp} \citep[e.g.][]{Khan+al:2011a,Khan+al:2011b}. They are short ($<$1\,min) in the low corona, but go up to a fraction of a day higher up. As the faint large-scale corona shows only slow variation these long exposure times still contain valuable information.

The key resources of \cusp\ are summarized in \tab{T:instr}. To estimate the data rate, a full spectro-polarimetric frame (3 images) was assumed to be downlinked every 5 minutes, resulting in 150\,kbit/s. Nonetheless, a fast digital processing electronics (implementing pattern recognition and event centroiding) is required to analyse the frames produced by the image sensor in real time.


To achieve the required stray light rejection levels, the required absolute pointing (pitch/yaw) of the spacecraft should be $<$5' and the pointing stability (pitch/yaw jitter) 1" in 15 minutes. 


The possibility of removing the polarization unit allows the normal polarimetric mode (\pol) as well as a non-polarimetric mode (\nop). In \nop\ mode \cusp\ will be able to operate in a similar fashion to UVCS/SOHO, but with superior spatial and temporal resolution, because of the increased aperture and throughput.









\subsection{Visible light and IR Coronagraph (\vircor)}    \label{S:instr.vircor}



\vircor\ consists of two channels that provide maps of the magnetic field vector, the structure, and the electron density in the corona above the limb. In the infrared (IR) channel  the full Stokes vector is recorded, i.e., linear and circular polarization. Using a tunable narrow-band filter, the spectral profiles are recorded. The visible light (VL) channel captures the K-corona with a tunable broad-band filter.

\begin{figure*}
\centerline{\includegraphics[width=\textwidth,bb=26 31 502 218,clip=true]{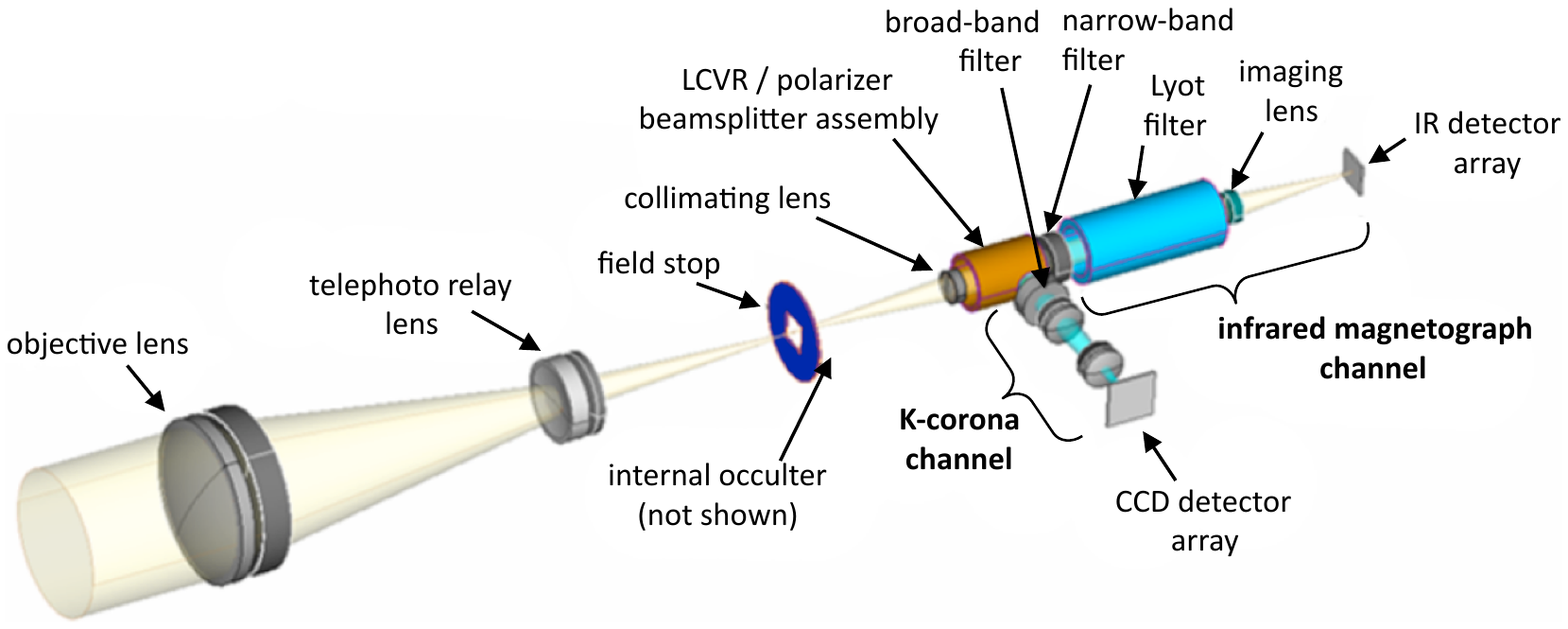}}
\caption{Schematic optical layout of the Visible light and IR Coronagraph (VIRCOR).
           \label{F:instr-vircor}}%
\end{figure*}


\vircor\ will be a complete instrument including mechanical structure, optical system, and local electronics. The Carbon Fiber Reinforced Plastic (CFRP) optical bench facesheets, baffles, and structural panels ensure that the stiffness and thermal stability requirements are met while achieving a low-mass structure.
The optical system is shown in \fig{F:instr-vircor}. The entrance aperture will be 20\,cm in diameter, large enough to achieve the required resolution and signal-to-noise ratio. The IR coronal magnetograph will fully profit from the unique observation conditions of \solmex\ in space because of the absence of seeing-induced polarization cross-talk and of atmosphere-induced intensity fluctuations. The large distance to the external occulter, located on the occulting spacecraft, enables high spatial resolution to be achieved in both the IR and the VL channel. 


The IR spectro-polarimeter
design is based on the HAO/NCAR Coronal Multi-channel Polarimeter
\citep[CoMP;][]{Tomczyk+al:2008}, which is now a part of HAO's Mauna Loa Solar
Observatory in Hawaii.
The 2.5\,$R_\odot$ $\times$ 2.5\,$R_\odot$ FOV (\fig{F:instr-fov}) is imaged onto a Teledyne Imaging 
HgCdTe 1024$\times$1024 pixels detector with 18\,$\mu$m square pixels (2.3"/pixel).
It records the intensity as well as the the linear and circular polarization of the forbidden lines of \ion{Fe}{13} at 1074.7 nm and 1079.8 nm, and of the \ion{He}{1} line at 1083\,nm. This provides information on the full vector of the coronal magnetic field.
The IR channel measures the line-of-sight plasma velocity from Doppler observations
and the plasma density from the ratio of the \ion{Fe}{13} lines at 1074.7\,nm and 1079.8\,nm. 
Liquid Crystal Variable Retarder technology is used for both the polarimetry analysis and tunable wavelength selection. Post-focus instrumentation includes a narrow-band tunable filter to obtain precise polarimetry across the emission lines over the entire FOV with modest spectral resolution. A six-stage birefringent filter will be used to attain the wavelength range.

The visible light (VL) channel has a high spatial sampling of 1.18"/pixel and a more extended 5\,$R_\odot$ $\times$ 5\,$R_\odot$ FOV imaged on a 4k $\times$ 4k CCD.
%
%
This provides context for the fine-scale features and takes full advantage of the high flux and excellent straylight rejection provided by the large external occulter distance. To satisfy the mission objectives, it provides total and polarized brightness images of the K-corona, i.e., of light scattered by free electrons in the corona. Using a broad-band filter ensures a high photon flux at the detector plane to allow short exposures down to 1\,s and the highest possible spatial resolution, surpassing ground-based eclipse imaging. Tuning the broad-band filter in wavelength provides access to temperature and flow diagnostics as required by the science objectives.


The key resources are summarized in \tab{T:instr}. A data rate of 300\,kbps is required to downlink the full Stokes parameters every 30\,s and the white light K-corona images every 5 minutes using compression.
%
%
To achieve the required stray light rejection levels, the absolute pointing of the spacecraft has to be ${<}1'$. The pointing stability must ensure that the same region on the Sun is imaged on each detector pixel within the sequence of spectro-polarimetric exposures.


\vircor\ will operate mainly in a synoptic mode. To investigate rapid variations, high-cadence observations will be performed for limited times. Then either only part of the FOV will be transmitted or the instrument will observe at low temporal cadence until the burst of high-cadence data is downlinked. To point to different parts of the corona, a spacecraft roll will have to be coordinated with the other instruments.

\subsection{EUV Imaging Polarimeter (\eip)}  \label{S:inst.eip}


\eip\ is a normal incidence EUV full-disk telescope that measures the linear polarization of the \ion{Fe}{10} line at 17.4\,nm in order to map the magnetic field orientation in the corona. To derive the Stokes $I$, $Q/I$ and $U/I$ parameters, the intensity is measured at three orientations (0$^\circ$, 60$^\circ$, 120$^\circ$) by rotating the focal plane assembly (FPA) that is composed of a polarizing mirror and the detector. The amplitude of the expected polarization signal is of the order of $10^{-3}$ (\sect{S:ondisk.corona}, \fig{F:FeX-polarization}).
A 3$\sigma$ detection of this signal requires about $10^7$ photons per exposure. The main design driver is therefore to collect enough photons within exposure times shorter than the typical evolution time scale of the observed structures.

\begin{figure*}
\includegraphics[width=\textwidth,bb=26 31 557 182,clip=true]{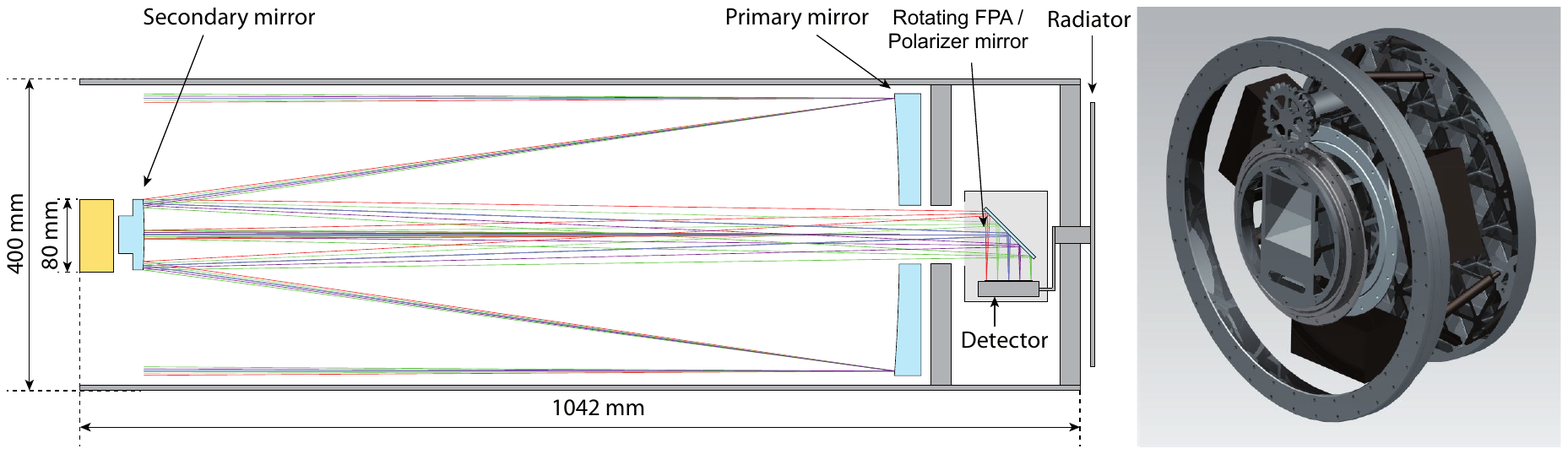}
\caption{Optical design of the EUV Imaging Polarimeter (EIP) and its rotating focal plane assembly (FPA).
~~~~~~~~~~~~~~~~~~~~~~~~~~~~~~~~~~~~~~~~~~~~~~~~ 
           \label{F:instr-eip}}%
\vspace*{-2ex}
\end{figure*}


\eip\ will be a Ritchey-Chr\'etien telescope (\fig{F:instr-eip}). The linear polarizer is a 45$^\circ$ folding mirror located close to the detector (${\approx}$10\,mm), and this focal plane assembly (FPA) is rotated as a whole to provide polarization measurements. This single mirror polarizing system maximizes the throughput, and ensures
a stable image on the detector, allowing us to measure Stokes $Q/I$ and $U/I$ independent of the flatfield. To keep the image stable within a pixel, the rotation axis of the mechanism must be maintained within 1 arcmin, which is achievable with careful design. The rotating FPA imposes that cooling lines and electrical connectors follow the movements of the detector.

The full disk (0.6$^\circ$x0.6$^\circ$) is imaged on a 4k\,x\,4k detector with 0.55"/pixel, providing a spatial resolution of 1.1".
The detector is a back-thinned EUV sensitive CCD with 12\,$\mu$m square pixels (e.g., E2V CCD203), or equivalent APS. In order to minimize the detector dark current during the exposure times, the array will be cooled to about $-80^\circ$\,C, so that the measurements are effectively shot-noise limited. 

\begin{figure}
\parbox[b]{0.55\textwidth}{\includegraphics[width=0.55\textwidth]{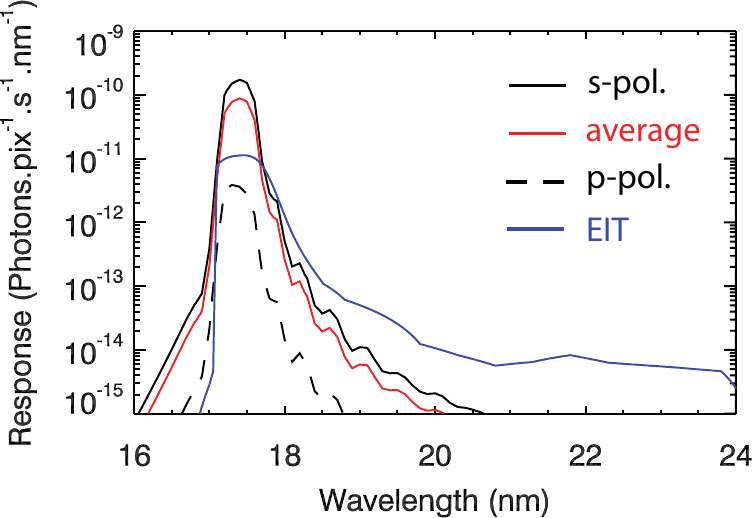}} \hfill
\parbox[b]{0.40\textwidth}{\caption{Spectral response of EIP for the s- and p-polarizations (solid and dashed) and their average (red), including the efficiencies of all optical elements.  The EIT/SOHO response (blue) is shown for comparison.
           \label{F:instr-eip-pol}
}}
\end{figure}

EIP uses state-of-the-art low-roughness Al/B$_4$C/Mo multilayer coatings with 50\% reflectivity. They are optimized for maximum throughput at 17.4\,nm and maximum rejection of nearby lines of \ion{Fe}{9}, {\sc x}, and {\sc xi}, some of which are predicted to show weaker or null polarization (e.g., \ion{Fe}{9} at 17.1\,nm). It is of critical importance  to reject these other lines because they constitute stray light in the polarization measurement and degrade the measurement signal-to-noise. 
We achieve the required high spectral purity by using the 2nd order of multilayers tuned to twice the 17.4\,nm wavelength, because the 2nd order has a narrower passband. The 1st order is suppressed by the Zr filter at the focal plane. The resulting passband is only ${\approx}0.35$\,nm wide (\fig{F:instr-eip-pol}) and a spectral analysis reveals a spectral pureness of about 70\%, which is twice as good as for traditional coatings.  
The thicknesses of the coatings can be controlled with a precision to ensure centering of the passband within 0.1\,nm.

Multilayer coatings are also used for the 45$^\circ$ polarizer. They are optimized to reflect only the s-polarization at 45$^\circ$ of incidence. The p-polarization is suppressed by about a factor of 100 (\fig{F:instr-eip-pol}). 
Nonetheless, the relative s-polarization can be measured with a high accuracy of $10^{-3}$ \citep{Toro-Iniesta+Collados:2000}, because a non-zero p-polarization is only reducing the amplitude of the response when rotating the polarizing mirror (with the whole FPA).

Thin film ($\approx$150\,nm) metallic filters are used at the entrance of the telescope and at the focal plane. The entrance Al filter suppresses the visible and infrared light. The focal plane filter is an Al/Zr/Al sandwich. It suppresses the 1st order of the multilayer coatings and provides redundancy against pinholes developing in the front filter during the mission lifetime.


We computed the expected signal at the focal plane for the quiet Sun and for active regions with the measured radiances and the polarization levels from \sect{S:ondisk.corona} and \fig{F:FeX-polarization}. From this we estimated the exposure times for \eip\ using a 2"x2" binning.
For an active region the required polarization level of $ 7{\cdot}10^{-4} $ can be reached within a 43\,s exposure (1$\sigma$ detection).
%
%
Because three exposures (at 0$^\circ$, 60$^\circ$, 120$^\circ$ polarizer angle) have to be taken, a magnetic field measurement can be performed in 3\,min at 1$\sigma$ detection level. Our spectral analysis shows that the other lines in the band contribute only 4\% to the polarization when compared to the EIP target line.


The key instrument parameters are given in \tab{T:instr}. The pointing requirements are derived from the need to keep the image stable on the detector within a fraction of a pixel over the typical exposure times.
If the spacecraft pointing stability does not meet the instrument requirements, actuators on the secondary mirror will be used to correct small jitter errors.


The observation programme will consist of repeated sequences of three polarization images at 0$^\circ$, 60$^\circ$, and 120$^\circ$. The best compromise between signal level, resolution (binning) and exposure times will be determined during commissioning. Keeping the polarizer at a fixed position, \eip\ can take data at full resolution with a cadence of 3\,s to study the fast evolution of coronal structures.









\subsection{Scanning UV Spectro-Polarimeter (\susp)}  \label{S:instr.susp}


\susp\ will measure linear and circular polarization in the UV range from 115\,nm to 155\,nm, i.e., the full Stokes vector, with the prime lines for magnetic field diagnostics being \Lya\ and \ion{C}{4} using the Hanle and the Zeeman effect (cf.\ \sect{S:TR}). Other lines in this wavelength range cover a wide temperature range from $10^4$\, to well above $10^6$\,K, allowing plasma diagnostics also in the hot parts of the atmosphere.

\begin{figure}
{\includegraphics[width=\textwidth,bb=16 48 542 310,clip=true]{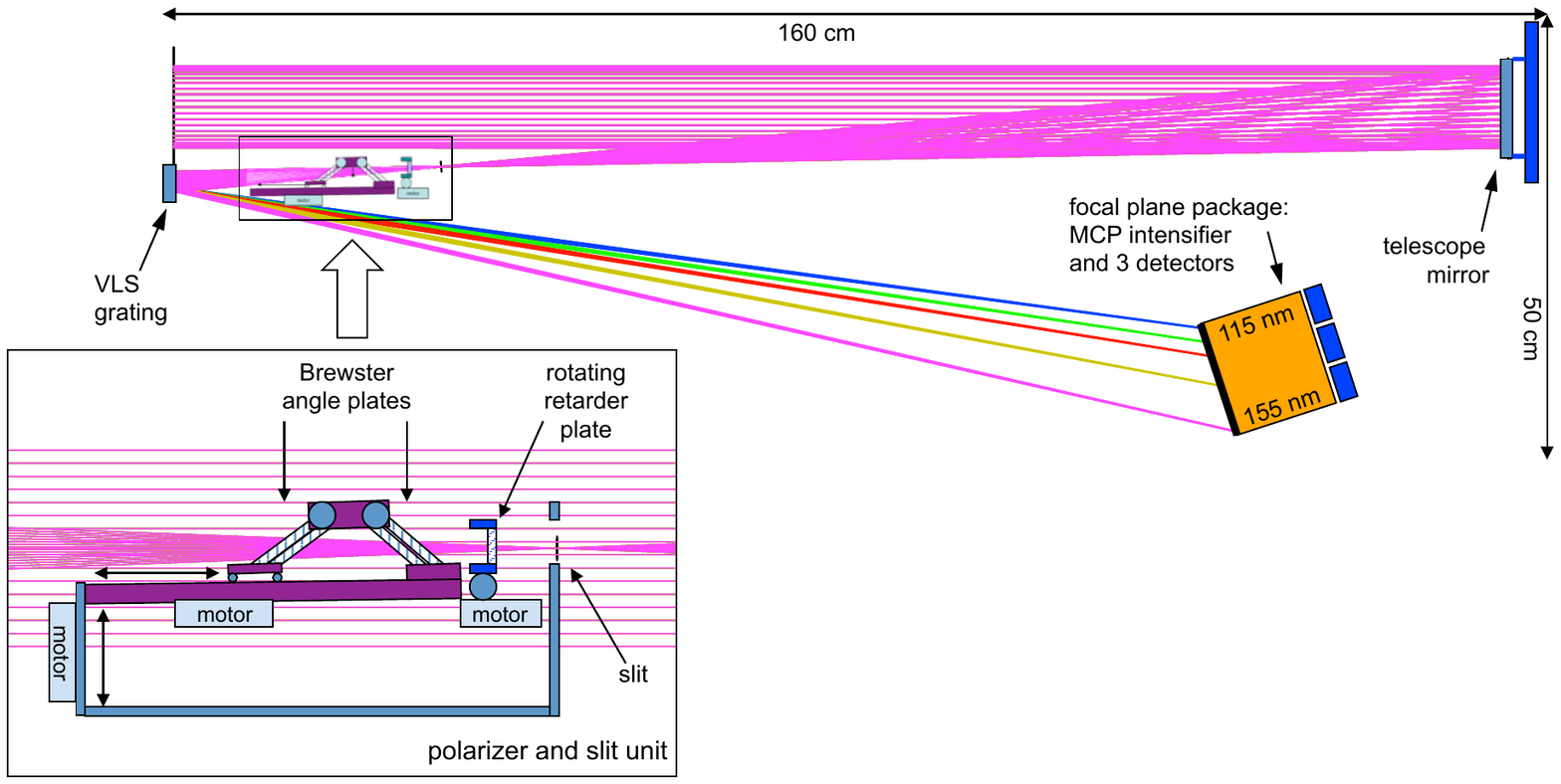}}
\caption{Scanning UV Spectro-Polarimeter (SUSP).
           \label{F:instr-susp}
A single mirror telescope and a concave grating spectrometer image part of the UV spectrum on a system of three detectors. The inset shows a magnification of the polarizer and slit unit --- rotated by 90$^\circ$ around the beam axis for clarity. The polarizer unit can be removed from the light beam for non-polarimetric observations.
}
\end{figure}


SUSP is an efficient normal incidence UV slit spectrometer with very high throughput. It consists of a telescope with a single parabolic mirror and a grating spectrograph with a concave variable line spacing (VLS) grating (\fig{F:instr-susp}). In the focal plane a large format multi-channel plate (MCP) covers the entire spectral and spatial FOV (120\,mm x 30\,mm). The image produced by the anode phosphorous coating of the visible-blind MCP intensifier is transferred by fiber optical couplers to three separate CMOS-APS sensors (2k\,x\,2k pixels) that are read out in photon-counting mode. The telescope mirror mechanism allows us to point anywhere on the solar disk and up to $1\,R_\odot$ above the limb, providing an overlap with the coronagraphic instruments \cusp\ and \vircor. The transmissive polarization optics are placed between the entrance slit and the grating. The design of the polarizing unit follows \cite{Winter+Ortjohann:1987}, who measured 14\% transmission for the
  complete unit. It consists of a rotating MgF$_2$ retarder plate and a double set of Brewster-angle MgF$_2$ plates, with the Brewster angle perpendicular to the dispersion plane (\fig{F:instr-susp}).


The sampling is 1'' per pixel in the spatial and 6.6\,pm per pixel in the spectral direction, the latter is sufficient to measure flows with 2\,km/s accuracy (by line centroiding). Good image quality can be achieved over 360'' along the slit. To reach the required polarimetric sensitivity of $10^{-3}$ (cf.\ \sect{S:TR}, \figs{F:Lya-polarization}, \ref{F:stokes.civ}), a signal-to-noise level of $10^{3}$ has to be achieved, i.e., at least $10^6$ photons have to be detected (Poisson statistics). This can be achieved with a photon-counting detector.
The photon budget for the whole instrument shows that for quiet Sun observations $10^6$ counts are reached in 7\,s for \Lya\ in one pixel and in about 6\,min for \ion{C}{4} with 2''x2'' binning. With such such a binning a polarization signal of more than 0.1\% can be expected (see \fig{F:stokes.civ}) Active region observations or larger binning would shorten the necessary exposure times.


The key resources of \susp\ are listed in \tab{T:instr}. The data rate was estimated assuming uninterrupted observations downlinking wavelength windows for selected lines and lossless compression. The pointing stability has to be sufficient to keep the slit on the same structure. Spacecraft absolute pointing must ensure to find the target well within the FOV provided by the raster scan.


\susp\ will have a polarimetric (\pol) and a non-polarimetric (\nop) observing mode. The polarization optics can be retracted from the optical path entirely (inset of \fig{F:instr-susp}).
In the {\sc{pol}} mode the instrument observes different polarization states, i.e., at different rotation angles of the retarder plate, in rapid succession. For longer exposure times, which are needed for \ion{C}{4}, the polarization states are then summed separately over the data sequence, e.g., 12 min $=$ 2\,x\,6\,min for quiet Sun \ion{C}{4}, and analysed to return magnetic field information. This is the fundamental reason why a photon counting detector is required. In the \nop\ mode the polarizer unit is removed from the light beam and the UV spectra can be acquired in unprecedented time cadence for plasma diagnostics. In the \pol\ and \nop\ modes either a time series of spectra or raster maps can be acquired.

\subsection{Chromospheric Magnetic Explorer (\chrome)}    \label{S:instr.chrome}


\chrome{} will provide chromospheric and photospheric vector magnetic field
maps using spectro-polarimetric measurements in the core of the chromospheric \mgiik{}
279.6\,nm line and a photospheric line. %
The combined action of atomic level polarization and the Hanle and Zeeman effects creates and modifies polarization signatures, allowing us to retrieve the direction and strength of the magnetic field.

\begin{figure*}
\includegraphics[width=\textwidth]{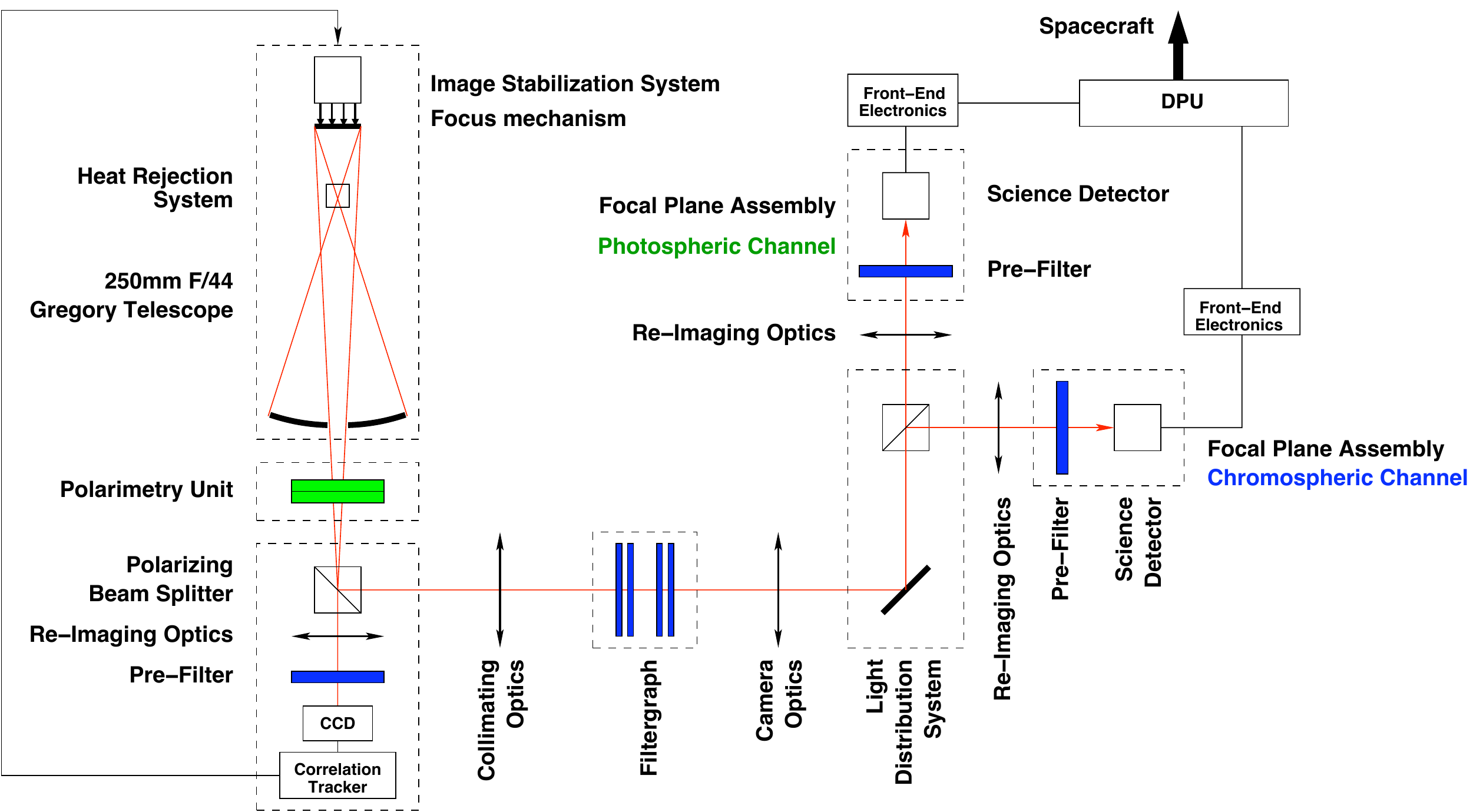}
\caption{Conceptual design of the Chromospheric Magnetic Explorer (\chrome).
           \label{F:instr-chrome}}
\end{figure*}


The conceptual design of \chrome{} is sketched in \fig{F:instr-chrome}.  The
25\,cm aperture Gregory-type telescope uses an active mirror (M2) for internal
image stabilization and refocussing.  The polarimetry unit is located
before any inclined optical element to minimize instrumental polarization.
The polarizing beam splitter acts as the ana\-lyser of the polarimetry
unit. At the same time it feeds light to the
correlation tracker sensor. The filtergraph consists of a double Fabry-P\'erot
interferometer, mounted in a collimated setup to keep a constant spectral
resolution of 50\,m\AA{} at 280\,nm over the whole FOV. Behind the
filtergraph a dichroic beam splitter distributes the light to the
chromospheric and the photospheric channels. The chromospheric channel is equipped with a $\approx$15\,\AA{} wide pre filter. The photospheric channel is
currently designed to use the \ion{Fe}{1} line at 525\,nm, as there is no sufficiently strong high-$g$ unblended line at shorter wavelengths.


\chrome{} delivers chromospheric vector magnetograms at a spatial resolution
of 0.30" (with 0.15"/pixel sampling) and a spectral resolution of 50\,m\AA{}. The FOV
is 300\arcsec{}$\times$300\arcsec{} and the nominal cadence is 30\,s. Photon
budget considerations, based on measurements of the solar irradiance \citep{Cebula+al:1996}, on theoretical calculations of the Zeeman signal of the \mgiik{}
line, and the scattering polarization
signal modified by the Hanle effect \citep{Trujillo-Bueno:2011} yield a noise level
for the polarimetric measurements of $3{\cdot}10^{-3}$ of the intensity, sufficient to fulfill
the requirement for measuring the chromospheric Hanle and Zeeman signals over
network and active regions. On-board spatial binning and/or longer exposure
times, resulting in a polarimetric precision of $10^{-4}$, allow for
chromospheric magnetic field measurements also in quiet Sun areas. 
The \ion{Fe}{1} line for the photospheric channel
permits reliable photospheric vector magnetograms and detailed magnetic
field investigations \citep[e.g.][]{Solanki+al:2010.sunrise}. The photospheric channel uses the same filtergraph as the
chromospheric channel. Simultaneous observations in the chromospheric and the
photospheric channel can be achieved by adjusting the free spectral range
(FSR) of the etalons accordingly.


\tab{T:instr} summarizes the baseline properties of \chrome{}.  The
full performance of \chrome{} is exploited if simultaneous measurement of the chromospheric and the photospheric magnetic field vector are acquired within 30\,s over the full FOV. This would result in a very high data rate of 15\,MByte/s. However, through an efficient on-board data selection (smaller FOV, lower cadence at times of low activity) and compression it is possible to reduce the maximum data rate to 700\,kBit/s.


The required pointing accuracy of \chrome{} is 1/20 pixel on the science
detector ($7.5\times 10^{-3}$\,arcsec). This will be achieved by an internal
image stabilization system (correlation tracker), and reduces the pointing
requirement for the spacecraft to 10\arcsec{} per minute.
Repointing of the FOV should allow for observations on the entire
solar disk.


The default operation mode of \chrome{} will deliver full Stokes images at 15
wavelength positions simultaneously in the \mgiik{} line and the photospheric
line. The 15\,\AA{} wide prefilter of the chromospheric channel allows for an
extended observing mode that additionally covers the non-Hanle-sensitive \ion{Mg}{2}\,h line, removing ambiguities in the magnetic field determination. A fast
observing mode using fewer wavelength points and/or only circular or linear
polarization states will be implemented to allow the investigation of wave
phenomena and reconnection events.

\section{Model spacecraft, mass and power\label{S:spacecraft}}


To obtain a feasible design for the spacecraft system and its components and to derive reliable estimates for the mass, power, and costs for the \solmex\ mission, we performed a concurrent engineering (CE) study at the DLR Institute of Space Systems in Bremen, Germany.
\nnn{While the basic idea for the formation flight is similar to previously proposed missions such as DynaMICCS \cite{Turck-Chieze:2009} and especially \compass\ \citep{Fineschi+al:2007}, the spacecraft concept and related issues such as telemetry or the formation flight concept have been newly developed from scratch during the CE study.} 
A short summary of the study report's results are provided in the following.

\begin{figure}
\includegraphics[width=\textwidth,bb=26 31 460 337,clip=true]{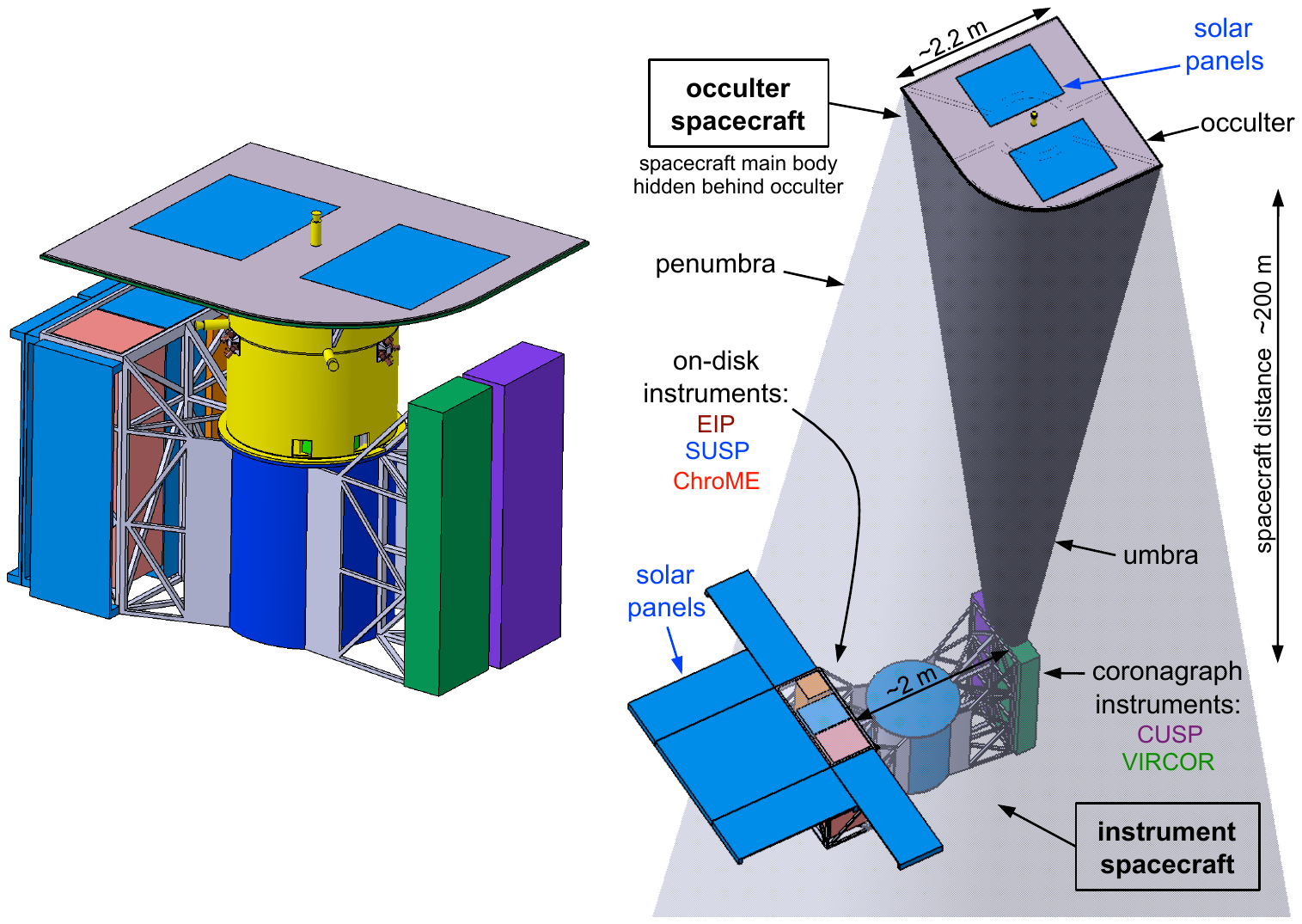}
\caption{%
\emph{The left panel} shows both spacecraft stacked on top of each other in the launch configuration. The circular main body of the instrument spacecraft (blue) is mounted on the circular launch adapter of the Soyuz Fregat. The occulter spacecraft also has a circular main body (yellow) and is mounted on the top.
\newline
\emph{The right panel} shows the flight configuration. The umbra of the occulter covers only the entrance apertures of the coronagraphic instruments, while the on-disk instruments and the solar panels are outside the penumbra. The Sun would be found towards the top. (Spacecraft distance not to scale).
\label{F:sc}
}
\end{figure}

The design of the spacecraft structure is driven by the requirement that the apertures of the  coronagraphic instruments (CUSP, VIRCOR) have to be in the umbra of the occulter, and that the on-disk instruments (EIP, SUSP, ChroME) have to be outside the penumbra. Based on the science requirements, the distance to the occulter should be 200\,m (\sect{S:mission}, \fig{F:occulter}). As a consequence the dimensions of the occulter have to be about 2.2\,m x 2.5\,m. The radius of the penumbra at the instrument spacecraft is then about 2\,m, which sets the separation between the coronagraphic and the on-disk instruments (cf.\ \fig{F:sc}).

\emph{The instrument spacecraft} is built around a circular main body with a diameter matching the circular-shaped launch adapter of the Soyuz Fregat. This carbon fibre tube main body will accommodate most spacecraft subsystem components. Exceptions are, e.g., the star trackers. 
The instruments are then mounted on the main body through a light-weight frame structure, which also holds the solar panels (\fig{F:sc}, right panel). 

\emph{The occulter spacecraft} is also built around a circular body fitting on top of the instrument spacecraft in order to have a compact and stable launch configuration (\fig{F:sc}, left panel). Here, the subsystems are again accommodated in the central tube. The occulter plate is mounted on top of the tube and carries at its Sun-directed side solar panels sufficient to power the occulter spacecraft.

\nnn{In the currently proposed design all instruments are housed in the instrument spacecraft: the instruments for on-disk (outside the penumbra) and for above-limb coronagraphic observations (inside the umbra; see \fig{F:sc}).  This has the advantage that the demands on the occulter spacecraft are quite relaxed, e.g., in terms of pointing stability or power consumption. For example it is sufficient to place the solar cells for the occulter spacecraft at the front side of the occulter, without the need for external panels (see \fig{F:sc}). This way the \solmex\ system is a combination of an instrument spacecraft with a complexity common for solar space observatories and a relatively simple occulter spacecraft.
}

\emph{The launch configuration} has maximum extensions of 3.1\,m and 2.2\,m in the hori\-zontal directions and 1.9\,m in height. Because the launch configuration is not rectangular (\fig{F:sc}, left panel), it easily fits into the Soyuz Fregat with its 3.86\,m diameter.

The occulter is not rectangular, but has one round edge to allow imaging observations very close to the limb with the coronagraphic imaging spectro-polarimeter \vircor. The slit instrument \cusp\ uses the linear part of the occulter. Both \vircor\ and \cusp\ will be able to observe simultaneously.

For observations of different parts of the corona both the occulter and the instrument spacecraft have to roll. The positioning of the instruments implies that the roll axis intersects the instrument spacecraft at its side (i.e., at the entrance aperture of the coronagraphs). Therefore, the instrument spacecraft has to roll around a point not identical to its centre of mass. This is achieved through propulsion, i.e., cold gas micro thrusters.


The estimates for the overall resources are based on the concurrent engineering (CE) study for the mission profile and spacecraft. They contain margins between 10\% and 20\% depending on the technology readiness level. The instrument masses (cf.\ \tab{T:instr}) contain similar margins. Additionally a system margin of 20\% was added to the total dry mass.
The total \nnn{launch} mass of the instrument spacecraft \nnn{(including launch adapter and propellant)} is about 1400\,kg including the launch adapter, while the occulter spacecraft is just below 700\,kg. Thus the total mass is below the limit of 2.1\,t for a Soyuz Fregat launch to L1.

In the framework of the CE study the power consumption of all subsystems was estimated for a number of operating modes. The \emph{science mode} with the acquisition of the science data through the payload instruments will be used most of the time. The \emph{maneuver mode} for changing the attitude, spacecraft rolls, and de-saturation of the reaction wheels will be activated about once a day for about 30 min. These modes drive the requirements for power consumption and, thus, the size of the solar cell elements. The \emph{safe mode} for vehicle rescue is the driver for the battery power. It is assumed that the the spacecraft can run in safe mode for 1.5 days on battery power.

The peak power required for the instrument S/C is about 800\,W (including spacecraft and instruments), the occulter S/C demands a peak power of 370\,W. Each S/C has its own power system, and the solar cells are designed assuming a degradation of 3.75\% per year. The required area of the solar cells is 5.5\,m$^2$ for the instrument S/C and 2.6\,m$^2$ for the occulter S/C.
\nnn{The communication is realized with a data downlink of 7 Mbit/s employing a 60\,W transmitter  using a 0.7\,m diameter dish antenna. This provides sufficient bandwidth to ensure the downlink of the science data with an average rate of 2 Mbit/s (cf.\ \tab{T:instr}).}

\section{Summary and conclusions}  \label{S:conclusions}

The information on the magnetic field of the outer solar atmosphere is encoded in the polarization of ultraviolet and infrared spectral lines \citep[e.g.,][]{Trujillo-Bueno+al:2005.cv}, and was not accessible to previous space instruments. Because the level of polarization is generally low, the required signal-to-noise ratio for the observations has to be as high as $10^4$ for some measurements. Thus, the instruments have to provide a combination of large aperture and high throughput. 

Previous solar space missions did not include the capability of polarimetric observations, or it turned out that the signal-to-noise ratio was not sufficient to provide reliable results (e.g., {\sc uvsp/smm}). \solmex\ will overcome these limitations and will provide the first measurements of the magnetic field in the upper solar atmosphere.

Space observatories such as SOHO \citep{Domingo+al:1995}, TRACE \citep{Handy+al:1999}, Hinode \citep{Kosugi+al:2007}, and SDO \citep{Pesnell+al:2011} have provided us with a new picture of the state of the plasma and its dynamics in the upper solar atmosphere, and Solar Orbiter will exceed this with increased spatial resolution from vantage points not explored before. While these observatories provided (or will provide, in the case of Solar Orbiter) information on the flux and energy of the photons originating in the solar atmosphere, e.g., spectral radiance, \solmex\ will for the first time also explore their state of polarization, giving access to the magnetic field. This will open a new window to investigate the interaction of the magnetic field with the plasma and to address the leading question of the nature of the upper solar atmosphere.

To reach this goal, \solmex\ will employ imaging polarimetric and spectro-polari\-metric instruments operating at extreme UV, UV and infrared wavelengths. These will observe the upper solar atmosphere of the Sun on the disk and above the limb. For the corona\-graphic instruments an external occulter will be mounted on a second spacecraft in formation flight with the spacecraft carrying the instruments.

Modern solar physics started with the first magnetic field measurement in sunspots by G.\,E.\,Hale in 1908. This revolution has remained incomplete, however, because the measurements of the magnetic field have been restricted mainly to the solar surface. \solmex\ will complete these achievements by providing the first comprehensive measurements of the magnetic field in the outer atmosphere of our Sun.




\def\aap    {A\hbox{\&}A }
\def\apj    {ApJ }
\def\nat    {Nature }
\def\solphys{Solar Phys. }
\def\sci    {Science }
\def\ssr    {Space Sci. Rev. }

{ \small

}

\end{document}